%% file: main.tex
\documentclass[twocolumn,apj,numberedappendix,twocolappendix,iop]{openjournal}

\usepackage[T1]{fontenc}
\usepackage{graphicx}
\usepackage{amsmath, amssymb, booktabs}
\usepackage{newtxtext,newtxmath}    
\usepackage[dvipsnames]{xcolor}  
\usepackage[colorlinks,linkcolor=blue,citecolor=blue,urlcolor=blue ]{hyperref}
\usepackage{amssymb}
\usepackage{dsfont}
\usepackage{enumitem}
\usepackage{hyperref}
\usepackage{multirow}
\usepackage{natbib}

\usepackage{xcolor}
\usepackage{textgreek}
\usepackage[utf8]{inputenc}
\usepackage[english]{babel}

\usepackage{hyperref}
\hypersetup{
    unicode, 
    colorlinks=true,
    linkcolor=linkcolor,
    citecolor=linkcolor,
    filecolor=linkcolor,
    urlcolor=linkcolor,
}
\usepackage{color,colortbl}
\definecolor{linkcolor}{rgb}{0.0,0.3,0.5}
\usepackage{tensind}
\tensordelimiter{?}
\DeclareGraphicsExtensions{.bmp,.png,.jpg,.pdf}
\usepackage{verbatim}
\usepackage[normalem]{ulem}
\usepackage{orcidlink}
\usepackage{soul}
\usepackage{fontawesome}
\usepackage{lineno}

\graphicspath{ {./figs/} }

\begin{document}
\title{Celia: Catalog Embeddings Learned from cosmologIcal simulAtions}

\shorttitle{Celia: Catalog Embeddings from Simulations}
\shortauthors{Zhang et al.}

\author{Xiaowen Zhang\orcidlink{0009-0008-0370-6021}}
\email{xiaowen4@andrew.cmu.edu}
\affiliation{McWilliams Center for Cosmology and Astrophysics, Department of Physics, Carnegie Mellon University, Pittsburgh, PA 15213, USA}

\author{Matthew Ho\orcidlink{0000-0003-3207-8868}}
\email{matthewho@ias.edu}
\affiliation{School of Natural Sciences, Institute for Advanced Study, 1 Einstein Drive, Princeton, NJ 08540, USA}

\author{Anshul Kumar}
\email{}
\affiliation{Heinz College of Information Systems and Public Policy, Carnegie Mellon University, Pittsburgh, PA 15213, USA}

\author{Tiziana Di Matteo\orcidlink{0000-0002-6462-5734}}
\email{}
\affiliation{McWilliams Center for Cosmology and Astrophysics, Department of Physics, Carnegie Mellon University, Pittsburgh, PA 15213, USA}

\author{Rupert A.C. Croft\orcidlink{0000-0003-0697-2583}}
\email{}
\affiliation{McWilliams Center for Cosmology and Astrophysics, Department of Physics, Carnegie Mellon University, Pittsburgh, PA 15213, USA}

\author{Barnab\'as P\'oczos\orcidlink{0000-0002-2898-4705}}
\email{}
\affiliation{Machine Learning Department, School of Computer Science, Carnegie Mellon University, Pittsburgh, PA 15213, USA}

\label{firstpage}
\begin{abstract}
Cosmological hydrodynamical simulations connect observed galaxies to cosmology and the physics of galaxy formation. Each simulation is expensive, and different codes model galaxy formation in different ways. We present Celia (Catalog Embeddings Learned from cosmologIcal simulAtions), a self-supervised foundation model that maps catalogs of subhalos and their galaxy properties to an embedding vector for the purposes of compression and organization of astrophysical and large-scale structure information. Celia is pretrained without parameter labels on the IllustrisTNG, SIMBA and Astrid suites of CAMELS and their N-body counterparts, by reconstructing masked subhalo properties and predicting the representations of a teacher network. Simple linear models fitted to the embeddings recover $\Omega_{\mathrm m}$ with $R^2=0.999$, and $\sigma_8$ and both supernova-feedback amplitudes with $R^2\geq0.97$, in all three suites. The embeddings translate well to out-of-distribution simulations such as Swift-EAGLE, a code that Celia never saw. A linear model fitted on the three training codes predicts its $\Omega_{\mathrm m}$ with a correlation coefficient $r=0.985$, and linear models fitted on 20 labeled simulations reach $r\geq0.94$ for $\Omega_{\mathrm m}$, $\sigma_8$ and a supernova-feedback amplitude. We decompose the variation among the simulations of each suite into principal components of the embedding, and simple power laws of the simulation parameters describe the leading components. Celia gives simulated galaxy catalogs a shared coordinate system for simulation-based inference, emulation, the comparison of galaxy-formation models, the detection of physical degeneracies and the discovery of new astrophysical probes.
\end{abstract}

\maketitle

\input{01-Introduction}
\input{02-Method}
\input{03-Results_01}
\input{04-Results_02}
\input{05-Discussion}
\input{06-Conclusion}

\section*{Acknowledgments}

This work was supported by the Keystone Astronomy and AI (KAAI) program, funded by the Simons Foundation through its Targeted Grants to Institutes program (SFI-MPS-T-Institutes-00021638). We are grateful for discussions with Francisco Villaescusa-Navarro and with participants of the KAAI program, including Yixi Zhao, Sean Lewis, Fatemeh Hafezianzadeh and Patrick LaChance. We also acknowledge funding from the Block Center for Technology and Society at Carnegie Mellon University. TDM and RACC acknowledge funding from NASA ATP 80NSSC18K101 and NASA ATP NNX17AK56G. TDM additionally acknowledges funding from NASA ATP 80NSSC20K0519, NSF PHY-2020295, and NASA Theory grant 80NSSC22K072.

The computational results presented have been achieved using Vera at the Pittsburgh Supercomputing Center (PSC). ChatGPT (OpenAI) and Claude (Anthropic) were used to help format certain captions and figures and to polish the language; all text generated by these tools was carefully reviewed, edited, and verified for accuracy by the authors, who assume full responsibility for the outcome.

\bibliographystyle{mnras}

\bibliography{references}

\begin{appendix}

\input{07-Appendix}

\end{appendix}

\end{document}

%% file: 01-Introduction.tex
\section{Introduction}
\label{sec:intro}

Catalogs of galaxy positions and properties are precise probes of both the growth of cosmic structure and the physics of galaxy formation. The positions of galaxies trace the matter distribution, and their masses, sizes and stellar populations reflect how gas cools and forms stars within dark-matter halos \citep{wechsler2018galaxyhalo}. Modern surveys deliver such catalogs over large cosmological volumes. Spectroscopic surveys map the three-dimensional distribution of millions of galaxies \citep{desi2026dr1,desi2025dr2bao}, and imaging surveys measure weak gravitational lensing and galaxy clustering \citep{abbott2018desy1,des2026y6,li2023hscy3,wright2025kidslegacy}. Wide-area imaging with Euclid and the Rubin Observatory extends the reach of these measurements \citep{euclid2025overview,ivezic2019lsst,prat2023txpipe}. Studies of nearby satellites and high-redshift galaxies probe the same galaxy-formation processes over different environments and epochs \citep{geha2017saga,eisenstein2026jades}.

These data constrain cosmology and galaxy formation, whose effects are entangled on nonlinear scales. There, stellar and active galactic nucleus (AGN) feedback redistribute the gas, changing both galaxy populations and the matter distribution measured by lensing \citep{vandaalen2011owls,chisari2019feedback}. The resulting suppression of the matter power spectrum depends on the feedback model \citep{schneider2015bcm,vandaalen2020powerspectra} and affects comparisons between lensing and cosmic microwave background constraints \citep{amon2022s8,mccarthy2023flamingo}. These scales also contain information beyond two-point statistics \citep{villaescusa2020quijote}. Simulation-based analyses now use nonlinear galaxy clustering and weak-lensing maps in survey data \citep{hahn2023simbig,jeffrey2025desy3}, making the treatment of baryonic physics an integral part of cosmological inference \citep{huang2019baryons,arico2023desy3}. Hydrodynamical simulations impose physical priors on this entanglement, which makes them valuable for understanding galaxy formation and for constraining cosmology.

Hydrodynamical simulations follow structure formation and galaxy evolution together \citep{vogelsberger2020review,crain2023review}. IllustrisTNG, EAGLE, SIMBA, Astrid and FLAMINGO implement star formation, black-hole growth and feedback through different subgrid prescriptions \citep{pillepich2018tng,pillepich2018first,schaye2015eagle,dave2019simba,bird2022astrid,ni2022astrid,schaye2023flamingo}. Selected parameters are calibrated against observables such as the galaxy stellar mass function and cluster gas fractions \citep{crain2015eagle,pillepich2018tng,kugel2023flamingo}. From the same initial conditions, different models predict different galaxy masses and structures \citep{scannapieco2012aquila}. Their predictions also differ for the matter power spectrum and for the response of dark matter to baryons \citep{vandaalen2020powerspectra,gebhardt2026backreaction}. CAMELS and DREAMS sample this model dependence by varying cosmological or dark-matter parameters together with feedback parameters across large simulation ensembles \citep{villaescusa2021camels,villaescusa2023release,rose2025dreams}. CAMELS repeats these variations with several galaxy-formation models \citep{ni2023camels}, providing catalogs in which the effects of cosmology and feedback can be studied across different physical implementations.

Parameter inference illustrates both the information available and the challenge of transferring it between models. Simulation-based inference connects simulations to data without requiring a tractable likelihood \citep{cranmer2020sbi,alsing2019delfi,ho2024ltuili}. Neural compression has been used to retain cosmological information while marginalizing over baryonic effects \citep{villaescusa2022marginalize}. Parameter information has been extracted from two-dimensional maps \citep{villaescusa2022cmd}, individual galaxies and small galaxy populations \citep{villaescusa2022onegalaxy,echeverri2023onegalaxy,chawak2024multiple,liao2026onegalaxy}, and luminosity functions and colors \citep{lovell2025luminosity}. Transfer between simulation models depends on the input properties. Positions and velocities have supported robust inference of $\Omega_{\mathrm m}$ across hydrodynamical codes \citep{desanti2023robust,shao2023halos}, and internal galaxy properties can introduce stronger model dependence \citep{villaescusa2022onegalaxy,echeverri2023onegalaxy,shao2023halos}. Domain adaptation, domain generalization and training across several models improve this transfer \citep{roncoli2023dagnn,lee2024domaingen,jo2025miest}.

Inference and emulation both need a description of an entire catalog, including its spatial arrangement, the properties of its members and the relations among them. Set and point-cloud networks provide such a description through pooling and attention \citep{zaheer2017deepsets,qi2017pointnet,vaswani2017transformer,lee2019settransformer,zhao2021pointtransformer}. Graph and point-cloud models already infer cosmological parameters from halo and galaxy catalogs \citep{villanueva2022cosmicgraphs,makinen2022cosmicgraph,anagnostidis2022pointnet,desanti2023robust,shao2023halos,chatterjee2025pointcloud,baldwin2026inductivebiases}. Generative models emulate catalogs, evaluate their likelihoods or populate dark-matter simulations with galaxies \citep{cuesta2024pointcloud,nguyen2026nehod,pandey2025galactification}. Benchmarks such as CosmoBench bring several of these tasks together \citep{huang2025cosmobench}. Each of these networks addresses one task with one set of inputs, and what it extracts from the catalogs serves that task alone. A representation that could be reused across these analyses would make more of the information in each simulation available without learning a new catalog encoder for every new application.

Pretraining separates representation learning from the downstream task \citep{bommasani2021foundation,brown2020gpt3}. Self-supervised objectives learn from the data themselves by reconstructing masked inputs \citep{devlin2019bert,he2022mae,pang2022pointmae}, predicting representations of masked regions \citep{assran2023ijepa}, matching augmented views \citep{grill2020byol}, or aligning paired modalities \citep{radford2021clip}. Such reuse is valuable for hydrodynamical simulations, whose computational cost limits their number, volume and diversity. Each CAMELS Latin-hypercube set used here contains 1,000 simulations of a $(25\,h^{-1}\,\mathrm{Mpc})^3$ box \citep{villaescusa2021camels}. Pretraining on N-body simulations has already reduced the number of hydrodynamical simulations needed for an inference task \citep{saoulis2025multifidelity}. In astronomy, pretrained image and spectrum representations support multiple measurements of galaxy properties \citep{hayat2021ssl,parker2024astroclip,parker2025aion}, and related approaches learn from simulated galaxies, cosmological fields and physical simulations \citep{xia2025multimodal,akhmetzhanova2024ssl,kannan2025cosmoflow,mccabe2025walrus,mukhopadhyay2026turbulence}. \citet{ting2026review} reviews these developments.

Shared representations can organize the differences between physical models. To measure the baryonic suppression of the matter power spectrum, \citet{lin2026onelatent} train a two-dimensional latent model jointly on the IllustrisTNG, Astrid and Swift-EAGLE suites. Pretrained representations also transfer between data sets and scientific tasks. Models of particle sets in collider physics can be reused across tasks \citep{golling2024mpm,mikuni2025omnilearn}, and OmniCosmos applies such a model to cosmological inference from halo and galaxy point clouds \citep{mikuni2026omnicosmos}. In 21-cm cosmology, SKATR learns summaries that support parameter inference and can be transferred, with a frozen encoder, to a different simulator \citep{ore2025skatr,pietschke2026crosssim}. For galaxy catalogs, the corresponding question is whether one representation can retain the responses to both cosmology and feedback across different galaxy-formation models, and make that information accessible to simple downstream readouts.

Here we present Celia, a self-supervised foundation model for simulated galaxy catalogs. We pretrain its encoder on the subhalo catalogs of the IllustrisTNG, SIMBA and Astrid suites of CAMELS and on their N-body counterparts, without simulation-parameter labels in the pretraining objective. The encoder learns the relations among subhalos and their properties by reconstructing the properties hidden from a masked view and by predicting the representations that a teacher network assigns to the same subhalos in the complete view. After pretraining, the frozen encoder produces a fixed-length summary of each catalog, and parameter inference requires only a linear readout. We test how well such readouts recover the cosmological and feedback parameters, how the summary is transferred to Swift-EAGLE, a code outside the training suites, and what the principal axes of the embedding reveal about each simulation suite. Section~\ref{sec:method} presents the simulations, encoder, training and readouts. Section~\ref{sec:results} reports the inference, transfer and representation analyses. Sections~\ref{sec:discussion} and~\ref{sec:conclusions} discuss the implications and summarize the conclusions.

%% file: 02-Method.tex
\section{Method}
\label{sec:method}
 
\begin{figure*}
 \centering
 \includegraphics[width=\textwidth]{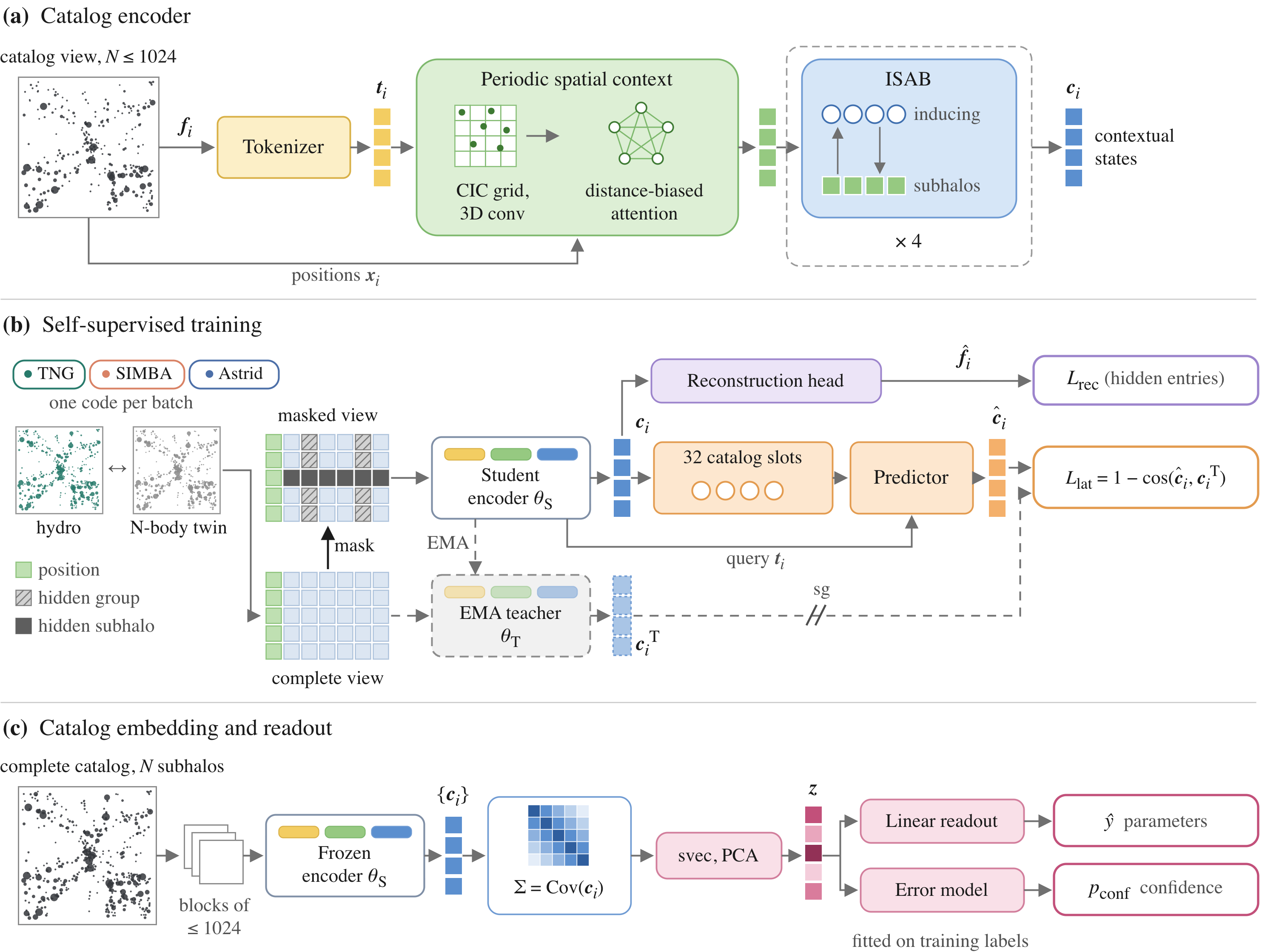}
 \caption{Overview of the method. (a) The encoder tokenizes each subhalo of a catalog view from its visible input features $\boldsymbol f_i$, adds periodic spatial context from a cloud-in-cell grid and from attention biased by the separations between subhalos, and contextualizes the tokens with four induced set-attention blocks (ISAB) into the contextual states $\boldsymbol c_i$. (b) Self-supervised training on hydrodynamical catalogs of one simulation code per batch and on their N-body counterparts. The student encoder sees a masked view, in which some input groups are hidden for all subhalos (hatched columns) and the members of randomly chosen host groups keep only their positions (dark rows). It reconstructs the hidden entries and uses 32 catalog slots to predict the states $\boldsymbol c_i^{\mathrm T}$ that the EMA teacher assigns to the partially hidden subhalos in the complete view. Dashed paths carry no gradient. (c) After training, the frozen encoder embeds complete catalogs in blocks of at most 1,024 subhalos. The covariance of the states is vectorized and projected onto principal components of the labeled catalogs to give the catalog embedding $\boldsymbol z$. A linear (ridge) readout predicts the parameters from $\boldsymbol z$, and an error model assigns each prediction a confidence score $p_{\mathrm{conf}}$.}
 \label{fig:method}
\end{figure*}
 
Figure~\ref{fig:method} outlines the method. We train an encoder (Section~\ref{sec:architecture}) on the subhalo catalogs of three CAMELS simulation suites (Sections~\ref{sec:data_simulations} and~\ref{sec:catalog_data}) with a self-supervised objective that at no point uses labeled simulation parameters (Section~\ref{sec:objectives}). After training, each catalog is summarized by the covariance of the object representations that the frozen encoder assigns to its subhalos (Section~\ref{sec:catalog_representation}). Linear readouts of this summary measure which parameters the representation encodes, and a second linear model assigns every prediction an error bar (Section~\ref{sec:readouts}). Appendices~\ref{app:details}--\ref{app:readout_details} give the input fields and the details of the encoder, its training and the readouts, and Appendix~\ref{app:pit} tests the calibration of the error bars.

\subsection{Simulations}
\label{sec:data_simulations}
 
We use the first generation of the Cosmology and Astrophysics with MachinE Learning Simulations \citep[CAMELS;][]{villaescusa2021camels,villaescusa2023release,ni2023camels}. Every hydrodynamical simulation follows $256^3$ dark-matter particles and initially $256^3$ gas particles from $z=127$ to $z=0$ in a periodic box of comoving side length $L=25\,h^{-1}\,\mathrm{Mpc}$. The cosmology is flat $\Lambda$CDM with $\Omega_{\mathrm b}=0.049$, $h=0.6711$, $n_{\mathrm s}=0.9624$, $w=-1$ and massless neutrinos. We use the snapshots at $z=0$.
 
The four suites differ in their hydrodynamics code and galaxy-formation model (Table~\ref{tab:data_suites}). The IllustrisTNG suite uses the moving-mesh code \textsc{arepo} \citep{springel2010arepo} with the IllustrisTNG model \citep{weinberger2017agn,pillepich2018tng}, and SIMBA uses the meshless finite-mass method of \textsc{gizmo} \citep{hopkins2015gizmo} with the SIMBA model \citep{dave2019simba}. Astrid uses the smoothed-particle hydrodynamics code MP-Gadget \citep{feng2018mpgadget} with the Astrid model \citep{bird2022astrid,ni2022astrid,ni2023camels}, and Swift-EAGLE uses the smoothed-particle hydrodynamics code \textsc{swift} \citep{schaller2024swift} with a galaxy-formation model based on EAGLE \citep{schaye2015eagle}. The models differ in their prescriptions for star formation, stellar feedback, black-hole growth and AGN feedback, and therefore populate halos of the same mass with different galaxies, gas and black holes.

Each suite has a Latin-hypercube (LH) set of 1,000 simulations that sample six parameters jointly, and every simulation has its own random seed for the initial conditions. Two cosmological parameters vary, $\Omega_{\mathrm m}\in[0.1,0.5]$ and $\sigma_8\in[0.6,1.0]$, together with four dimensionless amplitudes $A_{\mathrm{SN1}}$, $A_{\mathrm{SN2}}$, $A_{\mathrm{AGN1}}$ and $A_{\mathrm{AGN2}}$ that rescale the stellar and AGN feedback of each model relative to its fiducial value $A=1$. The feedback parameters carry the same names in all suites but scale different processes (Table~\ref{tab:data_suites}). $A_{\mathrm{AGN1}}$, for example, sets the energy of kinetic-mode AGN feedback in IllustrisTNG and Astrid, the momentum flux of quasar- and jet-mode feedback in SIMBA, and the Bondi accretion rate of black holes in Swift-EAGLE. $A_{\mathrm{SN1}}$ and $A_{\mathrm{AGN1}}$ vary within $[0.25,4]$ and $A_{\mathrm{SN2}}$ and $A_{\mathrm{AGN2}}$ within $[0.5,2]$, except $A_{\mathrm{AGN2}}$ in Astrid, which varies within $[0.25,4]$.
 
Sections~\ref{sec:results_response} and~\ref{sec:results_axes} use two further sets of IllustrisTNG, SIMBA and Astrid. In a one-parameter (1P) set, one parameter at a time takes its fiducial value and two values on either side of it, and all other parameters stay at their fiducial values. All 1P simulations share the random seed of their initial conditions. The cosmic-variance (CV) set keeps the fiducial parameters $\Omega_{\mathrm m}=0.3$, $\sigma_8=0.8$ and $A=1$, and its simulations differ only in the random seed of their initial conditions. For IllustrisTNG, we leave out the CV simulations that were run with a different code revision. The IllustrisTNG 1P simulations at the two values next to the fiducial one and at $\Omega_{\mathrm m}=0.1$ were run with other code revisions, and we keep them in all 1P analyses. The 1P and CV catalogs are embedded in the same way as the LH catalogs of their suite.
 
A fixed random permutation of the LH indices assigns 700 simulations of each suite to training, 100 to validation and 200 to testing, and every suite uses the same permutation. We used the validation catalogs to compare encoder configurations during development and to set the confidence tiers of Section~\ref{sec:results_pconf}. The readouts are fitted on the training catalogs, and the test catalogs enter only the evaluation. The self-supervised training uses the 2,100 training simulations of IllustrisTNG, SIMBA and Astrid and their 2,100 N-body counterparts. Swift-EAGLE is a held-out set. Its catalogs enter neither the training nor the normalization of the inputs. Its LH set is split in the same way and serves only to evaluate the frozen encoder (Section~\ref{sec:results_swift}).

\begin{table*}
 \centering
 \begin{tabular}{@{}lllll@{}}
  \toprule
  & IllustrisTNG & SIMBA & Astrid & Swift-EAGLE\\
  \midrule
  Code & \textsc{arepo} & \textsc{gizmo} & MP-Gadget & \textsc{swift}\\
  Hydrodynamics & Moving mesh & Meshless finite mass & SPH & SPH\\
  $A_{\mathrm{SN1}}$ & Wind energy per unit SFR & Wind mass loading & Wind energy per unit SFR & Energy per SN\,II event\\
  $A_{\mathrm{SN2}}$ & Wind speed & Wind speed & Wind speed & Metallicity dependence of\\
  & & & & stellar feedback energy\\
  $A_{\mathrm{AGN1}}$ & Kinetic-mode energy & Quasar- and jet-mode & Kinetic-mode energy & Bondi accretion rate\\
  & per unit accretion & momentum flux & per unit accretion & \\
  $A_{\mathrm{AGN2}}$ & Kinetic-mode ejection & Jet speed & Thermal-mode energy & Heating temperature\\
  & speed (burstiness) & & per unit accretion & of AGN feedback\\
  \midrule
  Role & Training, readouts & Training, readouts & Training, readouts & Readouts only\\
  LH catalogs & 700 / 100 / 200 & 700 / 100 / 200 & 700 / 100 / 200 & 700 / 100 / 200\\
  \bottomrule
 \end{tabular}
 \caption{The four CAMELS suites. Every suite varies $\Omega_{\mathrm m}\in[0.1,0.5]$, $\sigma_8\in[0.6,1.0]$ and four feedback amplitudes, which rescale the listed processes relative to the fiducial model ($A=1$). $A_{\mathrm{SN1}}$ and $A_{\mathrm{AGN1}}$ vary within $[0.25,4]$ and $A_{\mathrm{SN2}}$ and $A_{\mathrm{AGN2}}$ within $[0.5,2]$, except $A_{\mathrm{AGN2}}$ in Astrid, which varies within $[0.25,4]$. The last two rows give the role of each suite and the number of LH catalogs in the training, validation and test splits. SPH denotes smoothed-particle hydrodynamics, SFR the star-formation rate and SN\,II a Type II supernova.}
 \label{tab:data_suites}
\end{table*}

\subsection{Catalogs and input fields}
\label{sec:catalog_data}
\label{sec:data_catalogs}

Each input object is a gravitationally bound subhalo identified by \textsc{subfind} \citep{springel2001subfind,dolag2009subfind} in the CAMELS catalogs at $z=0$ \citep{villaescusa2023release}, together with the properties of its host friends-of-friends (FoF) group. The subhalo-to-group relation supplies the host environments, and the first subhalo of each group identifies its central. We keep centrals and satellites with dark-matter mass
\begin{equation}
 M_{\mathrm{DM}}\geq 50\,\frac{\Omega_{\mathrm m}^{\max}\rho_{\mathrm{crit},0}L^3}{256^3}
 =6.46\times10^9\,h^{-1}\,\mathrm M_\odot ,
 \label{eq:data_mass_cut}
\end{equation}
where $\Omega_{\mathrm m}^{\max}=0.5$ is the largest matter density of the hypercube and $\rho_{\mathrm{crit},0}=2.775\times10^{11}\,h^2\,\mathrm M_\odot\,\mathrm{Mpc}^{-3}$ is the critical density. The threshold is the mass of 50 particles in an N-body simulation with $\Omega_{\mathrm m}=\Omega_{\mathrm m}^{\max}$, so every selected subhalo contains at least 50 dark-matter particles in both the hydrodynamical and the N-body catalogs. The threshold is a fixed mass, so the selection does not change with cosmology through the particle mass. The same threshold applies to all suites.

The inputs use a subset of the fields of the \textsc{subfind} and FoF catalogs (Tables~\ref{tab:catalog_features} and~\ref{tab:catalog_features_host} in Appendix~\ref{app:details}). Besides its position, each subhalo is described by 30 properties of itself and of its host group and by a flag that marks whether it is the central subhalo of its group. Some properties are taken directly from catalog fields, such as the stellar mass or $V_{\max}$, and others are derived from them, such as the mass fractions of the host and the position and velocity of the subhalo relative to its host. A binary positivity indicator accompanies each of 13 properties that can vanish, such as the stellar mass and the star-formation rate, so that a measured zero differs from a missing value. A subhalo thus enters the encoder with 53 input features. Three of them are the coordinates of the position, and the properties give 36 continuous features, with three components for each vector property. The 13 positivity indicators and the central flag make up the remaining 14 features, which are binary. Masses, radii, rates and characteristic velocities such as $V_{\max}$ enter as logarithms, and all features except the positions are standardized with statistics of the training catalogs.
 
Every input feature carries an availability flag and a validity flag. A feature is unavailable when a catalog does not contain the underlying field, which is the case for the baryonic and black-hole properties of the N-body catalogs. It is invalid when its transformation is undefined, as for the logarithm of a vanishing mass, and the positivity indicator of such a mass remains valid. The encoder receives no label of the suite or of the kind of simulation. An N-body catalog differs from a hydrodynamical one only through its unavailable baryonic features and the properties of its subhalos.

The mass of a simulated subhalo is a multiple of the particle mass, which scales with $\Omega_{\mathrm m}$. We therefore omit the particle counts and perturb the base-10 logarithm $\ell_M$ of every mass that contains dark matter,
\begin{equation}
 \ell_M'=\ell_M+\operatorname{clip}(\epsilon,-0.05,0.05),\qquad
 \epsilon\sim\mathcal N(0,0.02^2),
 \label{eq:data_mass_perturbation}
\end{equation}
so that the encoder cannot read $\Omega_{\mathrm m}$ from the discreteness of the masses. Each offset is drawn once per object and stays the same in all training views and in the final embedding. The three host masses share one offset, and the host scales and fractions are computed from the perturbed masses.

\subsection{Encoder}
\label{sec:architecture}
 
The encoder takes a view of a catalog with up to 1,024 subhalos as input. It returns a contextual state $\boldsymbol c_i\in\mathbb R^{256}$ for each subhalo, and these states serve as the object representations (Fig.~\ref{fig:method}a). It builds a token for each subhalo from its visible features, adds spatial context from a periodic grid and from attention between all pairs of subhalos, and contextualizes the tokens with a Set Transformer \citep{lee2019settransformer}. All its operations are equivariant to permutations of the subhalos, so reordering the subhalos of a view reorders their states in the same way.

Let $\boldsymbol f_i\in\mathbb R^{53}$ collect the input features of subhalo $i$, and let $a_{ij}$, $v_{ij}$ and $o_{ij}$ be the availability, validity and visibility of feature $j$, where $o_{ij}=0$ marks an entry that is unavailable, invalid or hidden during training (Section~\ref{sec:training_masks}). An entry $f_{ij}$ is readable when $r_{ij}=a_{ij}v_{ij}o_{ij}=1$. The features other than the position form 21 input groups of closely related features, such as the three components of the velocity or a property and its positivity indicator (Appendix~\ref{app:groups}). For group $g$, the vector $\boldsymbol s_{ig}=|g|^{-1}\sum_{j\in g}(a_{ij},v_{ij},o_{ij},r_{ij})$ holds the fractions of its features that are available, valid, visible and readable, and each subhalo is first embedded as
\begin{equation}
 \boldsymbol u_i=\boldsymbol e_0+
 \sum_{g=1}^{21}
 \left[W_g(\boldsymbol r_{ig}\odot
 \boldsymbol f_{ig})
 +S_g\boldsymbol s_{ig}+\boldsymbol b_g\right],
 \label{eq:tokenization}
\end{equation}
where $W_g\in\mathbb R^{256\times|g|}$ projects the values of group $g$ without bias, $S_g\in\mathbb R^{256\times4}$ embeds $\boldsymbol s_{ig}$, $\boldsymbol b_g$ is a learned bias and $\boldsymbol e_0$ a learned base token. Unreadable values are set to zero before the projection, so a hidden or missing value reaches neither the token nor its gradient. The token $\boldsymbol t_i\in\mathbb R^{256}$ is obtained from $\boldsymbol u_i$ by layer normalization followed by a residual two-layer multilayer perceptron (MLP) with GELU activation.

Two modules add spatial context at different resolutions. The grid module supplies aggregated context, the environment of each subhalo on the scale of a grid cell ($1.6\,h^{-1}\,\mathrm{Mpc}$) and above. The attention module supplies pairwise context from the exact separations between subhalos, which resolve satellites and neighbors below this scale. In the first, a linear map projects each layer-normalized token to 32 channels $\boldsymbol h_i$, which are deposited with cloud-in-cell weights $w_{ik}$ onto the cells $k$ of a periodic $16^3$ grid. Each cell holds the weighted mean of the deposited channels and an occupancy term,
\begin{equation}
 \boldsymbol g_k=\left(\frac{\sum_i w_{ik}\boldsymbol h_i}{W_k},\ \ln(1+W_k)\right),
 \qquad W_k=\sum_i w_{ik}.
 \label{eq:cic_grid}
\end{equation}
Three-dimensional convolutions with circular padding process the grid, and trilinear interpolation with the same weights $w_{ik}$ returns the processed grid to the subhalos, where it is added to the tokens.
 
The second module is one layer of dense attention between all pairs of subhalos, biased by their separation. For fractional positions $\boldsymbol x_i$,
\begin{equation}
 \boldsymbol\delta_{ij}=
 \boldsymbol x_i-\boldsymbol x_j-
 \operatorname{round}(\boldsymbol x_i-\boldsymbol x_j),
 \qquad \rho_{ij}=\|\boldsymbol\delta_{ij}\|_2,
 \label{eq:periodic_separation}
\end{equation}
so that $\rho_{ij}$ is the minimum-image separation in units of $L$. For attention head $h$, the query--key score is
\begin{equation}
 \begin{split}
 e_{ij}^{(h)}={}&
 \frac{(\boldsymbol Q_i^{(h)})^{\mathsf T}
 \boldsymbol K_j^{(h)}}{\sqrt{32}}\\
 &+B_h\!\left(\rho_{ij},
 \ln[1+L_*\rho_{ij}],\ln[1+n_{\mathrm p}\rho_{ij}]\right),
 \end{split}
 \label{eq:spatial_attention}
\end{equation}
where $L_*\equiv L/(1\,h^{-1}\,\mathrm{Mpc})=25$ and $n_{\mathrm p}=256$, so that the arguments express the separation in units of the box, of $1\,h^{-1}\,\mathrm{Mpc}$ and of the initial mean particle spacing. A small MLP shared by the heads returns the biases $B_h$. Each spatial module is followed by a residual feed-forward block (Appendix~\ref{app:spatial}).

Four induced set-attention blocks \citep{lee2019settransformer} then contextualize the tokens. The tokens of the $N_{\mathrm v}\leq1{,}024$ subhalos of a view, after both spatial modules, form the rows of the state matrix $\mathsf H\in\mathbb R^{N_{\mathrm v}\times256}$, and each block updates it as
\begin{equation}
 \mathsf H\leftarrow\operatorname{MAB}\bigl(\mathsf H,\operatorname{MAB}(\mathsf I,\mathsf H)\bigr),
 \label{eq:isab}
\end{equation}
where $\mathsf I\in\mathbb R^{64\times256}$ holds 64 learned inducing vectors and $\operatorname{MAB}(\mathsf X,\mathsf Y)$ is a multi-head attention block in which the rows of $\mathsf X$ attend to those of $\mathsf Y$, so that its output has the shape of $\mathsf X$. The inducing vectors gather information from all subhalos of the view and return it to each of them. A final layer normalization of the rows of $\mathsf H$ gives the contextual states $\boldsymbol c_i\in\mathbb R^{256}$.

\subsection{Self-supervised training}
\label{sec:objectives}
\label{sec:training}
 
The encoder is trained as a student that sees a masked view of a catalog (Fig.~\ref{fig:method}b). The student reconstructs the hidden properties of each subhalo and predicts the representations that a teacher network assigns to the partially hidden subhalos in the complete view. Predicting the teacher's representations of hidden content follows the joint-embedding predictive architecture of \citet{assran2023ijepa}. The target state is given by a slowly updated teacher to keep the representation from collapsing to a constant.

\subsubsection{Views and masking}
\label{sec:training_masks}
 
For every training catalog, we draw 64 views of up to 1,024 subhalos from those above the mass threshold, and successive epochs cycle through them. Each view is rotated by one of the 24 rotations of the cube (the combination of $90^\circ$ rotations around three box axes), translated by a uniform periodic shift and permuted. The rotations act on the positions and on all vector fields. The student input, the teacher input and the reconstruction targets all use the same transformed view.

The student input then loses part of its fields randomly via grouped feature masking. Four fields are never drawn for masking. The positions, the host-relative positions and the central flag describe the set and its host structure, and the host dark-matter fraction is nearly fixed by the other host fractions. The positions are never hidden, and the other three fields are hidden only with their whole host in step~3 or by the rules of step~4. All other fields are masked in \emph{families}. A family consists of an input group and the groups that describe the same component or are computed from it. The velocity and the velocity relative to the host form one family, and the black-hole mass and the black-hole accretion rate form another. The three host masses $M_{\mathrm{FoF}}$, $M_{200\mathrm c}$ and $M_{500\mathrm c}$ also form one family. Of the 11 families, the four baryonic ones are built around the stellar mass, the gas mass, the star-formation rate and the black-hole mass (Appendix~\ref{app:groups}). The mask of a view is drawn in four steps.
\begin{enumerate}[leftmargin=*,itemsep=1pt]
 \item With probability 0.25, the view of a hydrodynamical catalog loses all four baryonic families. The student then sees an N-body-like description of the catalog, and the hidden baryonic fields remain reconstruction targets.
 \item Each remaining family is hidden independently for all subhalos of the view, with probability 0.5 for a non-baryonic family and $1/3$ for a baryonic family. Together with step~1, every family is hidden with probability 0.5, since $0.25+0.75\times1/3=0.5$.
 \item Each host group is chosen with probability 0.2, and all its members in the view lose all their fields except the positions. Hiding whole hosts keeps the host fields of a hidden subhalo from being read from the other members of its host.
 \item Two rules hide fields that could be computed from visible ones. The host dark-matter fraction is hidden whenever the host gas, stellar or black-hole fraction is hidden. Whenever the host masses are hidden, the host-relative positions and velocities, the host center offset and the central flag are hidden for all members of the host, since two members would otherwise give $R_{200\mathrm c}$ or $V_{200\mathrm c}$ and hence $M_{200\mathrm c}$ (Appendix~\ref{app:masking}). This rule hides the host-relative velocities without their family, so the velocities stay visible.
\end{enumerate}
The subhalo total mass equals the sum of the dark-matter, gas, stellar and black-hole masses. It is not an input, so that the baryonic mass stays hidden whenever the baryonic families are hidden. In the student input, a hidden entry has its availability, validity and visibility flags set to zero. A field missing from the catalog carries the same flags, so the student cannot tell which entries were hidden. Teacher input and reconstruction target selection use the original flags, and the teacher sees every available entry of the same view.

\subsubsection{Objectives and optimization}
 
The reconstruction head is a two-layer MLP with GELU activation. It maps each student state $\boldsymbol c_i$ to predictions $\widehat{\boldsymbol f}_i$ of the continuous features and to logits for the positivity indicators. The targets are all available and valid entries hidden from the student, except the host-relative positions and the central flag. They include the baryonic features hidden in step~1, so the encoder learns to predict the baryonic properties of a subhalo from its dark-matter properties and its environment. Binary targets that a visible indicator already determines are not scored (Appendix~\ref{app:reconstruction}).
 
A continuous target is scored in standardized units with the smooth-$L_1$ penalty of its residual $\Delta=\widehat f_{bij}-f_{bij}$, where $j$ indexes the feature, $i$ the subhalo and $b$ the view in the batch. The penalty is 
\begin{equation}
 H(\Delta)=
 \begin{cases}
  \Delta^2/2, & |\Delta|<1,\\
  |\Delta|-1/2, & |\Delta|\geq1.
 \end{cases}
\end{equation}
The 13 positivity indicators are scored with the binary cross-entropy $\ell_{\mathrm{BCE}}$ between the predicted logit and the original indicator value of 0 or 1. With $\ell_j$ the penalty of feature $j$, the loss averages first over the scored entries $\mathcal I_j$ of feature $j$ in the batch, then over the scored features $\mathcal J_g$ of a loss group $g$, and finally over the loss groups $\mathcal G_+$ with scored entries,
\begin{equation}
 \mathcal L_{\mathrm{rec}}=
 \frac{1}{|\mathcal G_+|}
 \sum_{g\in\mathcal G_+}\frac{1}{|\mathcal J_g|}
 \sum_{j\in\mathcal J_g}\frac{1}{|\mathcal I_j|}
 \sum_{(b,i)\in\mathcal I_j}
 \ell_j\!\left(\widehat f_{bij},f_{bij}\right).
 \label{eq:masked_loss}
\end{equation}
The targets fall into 20 loss groups of physically related features, so that groups with many features do not dominate the loss.

The teacher is a copy of the encoder whose weights follow an exponential moving average (EMA) of the student weights, $\theta_{\mathrm T}\leftarrow m\,\theta_{\mathrm T}+(1-m)\,\theta_{\mathrm S}$, and receive no gradient \citep{grill2020byol,assran2023ijepa}. It encodes the complete view and returns the states $\boldsymbol c_i^{\mathrm T}$. The student predicts these states from a masked view in two attention steps. First, 32 learned query vectors attend to embeddings of the student states $\boldsymbol c_i$ and then to one another. These catalog slots summarize the masked view in 32 vectors. Second, each subhalo $i\in\mathcal R$ reads its prediction from the slots, where $\mathcal R$ holds the subhalos that keep at least one visible group other than the position and have at least one hidden group. The query is the token $\boldsymbol t_i$ of the subhalo, which is built from its visible fields. It attends with eight heads to the slots, and a final layer normalization of the result gives the prediction $\widehat{\boldsymbol c}_i$. The loss is the mean cosine distance
\begin{equation}
 \mathcal L_{\mathrm{lat}}=\frac{1}{|\mathcal R|}\sum_{i\in\mathcal R}
 \left(1-\frac{\widehat{\boldsymbol c}_i\cdot\boldsymbol c_i^{\mathrm T}}
 {\|\widehat{\boldsymbol c}_i\|\,\|\boldsymbol c_i^{\mathrm T}\|}\right),
 \label{eq:latent_loss}
\end{equation}
with the teacher states held fixed. Subhalos that keep nothing but their positions enter only the reconstruction loss. The total loss is
\begin{equation}
 \mathcal L=\mathcal L_{\mathrm{rec}}+\lambda\,\mathcal L_{\mathrm{lat}},
 \qquad \lambda=0.3 .
 \label{eq:total_loss}
\end{equation}
This value is the ratio of the gradient norms of $\mathcal L_{\mathrm{rec}}$ and $\mathcal L_{\mathrm{lat}}$ with respect to the encoder weights at the start of training, so that the two terms contribute gradients of similar size.

Each batch holds 32 training catalogs of one hydrodynamical suite together with their N-body counterparts, and successive batches cycle through the three suites. We optimize the student with AdamW \citep{loshchilov2017adamw} for 18,000 steps with a warmup--stable--decay learning-rate schedule, and the EMA momentum $m$ rises from 0.996 to 1 during the decay (Appendix~\ref{app:details}). All results use the student weights at the end of this single training run.

\subsection{Catalog representation}
\label{sec:catalog_representation}
 
After training, we freeze the student model and make every available field visible (Fig.~\ref{fig:method}c). To encode a full catalog, we partition its $N$ subhalos above the mass threshold by a fixed random permutation into blocks of at most 1,024 and encode each block as one view. A last block with fewer than 1,024 subhalos is filled up with other subhalos of the catalog, which serve as context, and only the outputs of its own subhalos are kept. Each subhalo thus contributes one state $\boldsymbol c_i$.
 
We summarize the catalog by the centered covariance of its states,
\begin{align}
 \boldsymbol\mu&=\frac{1}{N}\sum_{i=1}^N\boldsymbol c_i,\\
 \Sigma&=\frac{1}{N}\sum_{i=1}^N
 (\boldsymbol c_i-\boldsymbol\mu)
 (\boldsymbol c_i-\boldsymbol\mu)^{\mathsf T}.
 \label{eq:catalog_covariance}
\end{align}
As an average over the subhalos, $\boldsymbol \mu$ and $\Sigma$ do not depend on the order of the subhalos and are normalized by their count. The symmetric vectorization $\boldsymbol\xi=\operatorname{svec}(\Sigma)\in\mathbb R^{32{,}896}$ keeps the diagonal and multiplies the upper off-diagonal entries by $\sqrt2$, so that it preserves the Frobenius inner product of covariance matrices.

Every readout is fitted on a set $\mathcal T$ of labeled catalogs. On these catalogs alone, we center the vectors $\boldsymbol\xi$ and compute their principal components, keeping all those with eigenvalues above $10^{-12}$ of the largest. With the mean $\overline{\boldsymbol\xi}_{\mathcal T}$ and the matrix $P_{\mathcal T}$ of retained components, the catalog embedding is
\begin{equation}
 \boldsymbol z=s_{\mathcal T}\,P_{\mathcal T}^{\mathsf T}
 (\boldsymbol\xi-\overline{\boldsymbol\xi}_{\mathcal T}),
 \label{eq:catalog_pca}
\end{equation}
where the scalar $s_{\mathcal T}$ sets the mean variance of the retained components over $\mathcal T$ to one, so that the regularization strength below is in units of this variance (Appendix~\ref{app:readout_details}).

\subsection{Linear readouts and error bars}
\label{sec:readouts}
\label{sec:pconf}
 
The readout targets are
\begin{equation}
 \begin{split}
 \boldsymbol y=\bigl(&\Omega_{\mathrm m},\sigma_8,
 \log_{10}A_{\mathrm{SN1}},\log_{10}A_{\mathrm{SN2}},\\
 &\log_{10}A_{\mathrm{AGN1}},\log_{10}A_{\mathrm{AGN2}}\bigr).
 \end{split}
 \label{eq:readout_targets}
\end{equation}
For each target $y$, the readout minimizes
\begin{equation}
 \sum_{n\in\mathcal T}
 (y_n-b-\boldsymbol w^{\mathsf T}\boldsymbol z_n)^2
 +\alpha\|\boldsymbol w\|_2^2
 \label{eq:ridge}
\end{equation}
over the weights $\boldsymbol w$ and the intercept $b$. A local readout is fitted on the 700 training catalogs of one suite, and a pooled readout on the 2,100 training catalogs of IllustrisTNG, SIMBA and Astrid together, with its predictions evaluated separately for each suite. The regularization strength $\alpha$ is chosen by a three-fold cross-validation on the training catalogs (Appendix~\ref{app:readout_details}). Over the test catalogs of a suite, we report the coefficient of determination $R^2=1-\sum_n(y_n-\widehat y_n)^2/\sum_n(y_n-\bar y)^2$, the root-mean-square error and the Pearson correlation $r$.

Every prediction $\widehat y(\boldsymbol z)$ comes with an error bar $\widehat y(\boldsymbol z)\pm q_\beta\,\widehat\sigma(\boldsymbol z)$ for a nominal coverage $\beta$ and with a confidence score $p_{\mathrm{conf}}$. Both require a model $\widehat\sigma(\boldsymbol z)$ of the error size and multipliers $q_\beta$ that calibrate the error bars to the coverage $\beta$. Like the readout, they are fitted on the training catalogs and then applied unchanged to the test catalogs.

The size of the readout error varies from catalog to catalog. A catalog with few massive black holes, for example, constrains the AGN feedback less than one with many. The error model therefore predicts the error size of each catalog from its embedding.

The error model is fitted to the out-of-fold residuals $r_n=y_n-\widehat y^{\,\mathrm{oof}}_n$, in which every training catalog is predicted by a readout that did not see its label, so that the residuals represent the errors on new catalogs. We write the error as $r=\sigma(\boldsymbol z)\,\varepsilon$, with $\varepsilon$ independent of $\boldsymbol z$. The variable
\begin{equation}
 Z_n=\ln\!\left(r_n^2+\eta\right),\qquad \eta=10^{-4}\,V_0,
 \label{eq:logsq}
\end{equation}
satisfies $\mathbb E[Z\mid\boldsymbol z]\simeq\ln\sigma^2(\boldsymbol z)+\kappa$, with the constant $\kappa=\mathbb E[\ln\varepsilon^2]$ \citep{harvey1976heteroscedastic}. Here $V_0$ is the variance of $y$ over the training catalogs, and the floor $\eta$ avoids $\ln0$. A ridge regression of $Z_n$ on $\boldsymbol z_n$ gives $\widehat Z(\boldsymbol z)$ and the error size $\widehat\sigma(\boldsymbol z)=\exp[\widehat Z(\boldsymbol z)/2]$, which is proportional to $\sigma(\boldsymbol z)$. Its regularization strength is chosen on the same grid and folds as for the readout.

The multipliers are computed out of fold as well. For each fold, an error model fitted to the residuals of a cross-validation within the other two folds predicts $\widehat\sigma^{\,\mathrm{oof}}_n$ for the catalogs of the held-out fold. With the normalized residuals $R_n=|r_n|/\widehat\sigma^{\,\mathrm{oof}}_n$ of the $N_{\mathcal T}$ catalogs in $\mathcal T$ sorted as $R_{(1)}\leq\dots\leq R_{(N_{\mathcal T})}$,
\begin{equation}
 q_\beta=R_{(\lceil (N_{\mathcal T}+1)\beta\rceil)} .
 \label{eq:conformal_q}
\end{equation}
This quantile follows split-conformal and cross-validation-based prediction intervals \citep{lei2018conformal,barber2021jackknife}. Applied to the readout fitted on all training catalogs, it calibrates the coverage empirically, and Appendix~\ref{app:pit} tests this calibration on the test catalogs. The unknown factor $e^{\kappa/2}$ in $\widehat\sigma$ cancels in the product $q_\beta\widehat\sigma$, so the distribution of $\varepsilon$ need not be known.

The confidence score compares the 68\% error bar with the spread of the parameter over the training catalogs,
\begin{equation}
 p_{\mathrm{conf}}(\boldsymbol z)=\max\!\left[0,\;1-\frac{\bigl(q_{0.68}\,
 \widehat\sigma(\boldsymbol z)\bigr)^2}{V_0}\right].
 \label{eq:pconf}
\end{equation}
For Gaussian errors, the 68\% half-width is close to $\sigma$, so $p_{\mathrm{conf}}\simeq1-\sigma^2(\boldsymbol z)/V_0$ is a per-catalog analogue of $R^2$, and its mean over a set of catalogs approximates $1-\mathrm{MSE}/V_0$.

%% file: 03-Results_01.tex
\section{Results}
\label{sec:results}
 
We now examine what the catalog embedding $\boldsymbol z$ contains. Unless stated otherwise, all results use the frozen encoder and the 200 test catalogs of each suite. We first visualize the latent space (Section~\ref{sec:results_pca}) and measure the parameter information that a linear map extracts from it, with all input features (Section~\ref{sec:results_readout}) and with the restricted inputs of earlier supervised studies (Section~\ref{sec:results_restricted}). We then apply the encoder to a simulation code that it never saw (Section~\ref{sec:results_swift}) and relate the geometry of the embedding to the physical effect of each parameter (Section~\ref{sec:results_response}). Section~\ref{sec:results_pconf} connects the confidence of the readout with the black-hole population of each catalog, and Section~\ref{sec:results_axes} follows the numbers of massive galaxies and black holes along the principal axes of the embedding.

\subsection{Visualization of the latent space}
\label{sec:results_pca}
 
\begin{figure*}[t]
 \centering
 \includegraphics[width=\textwidth]{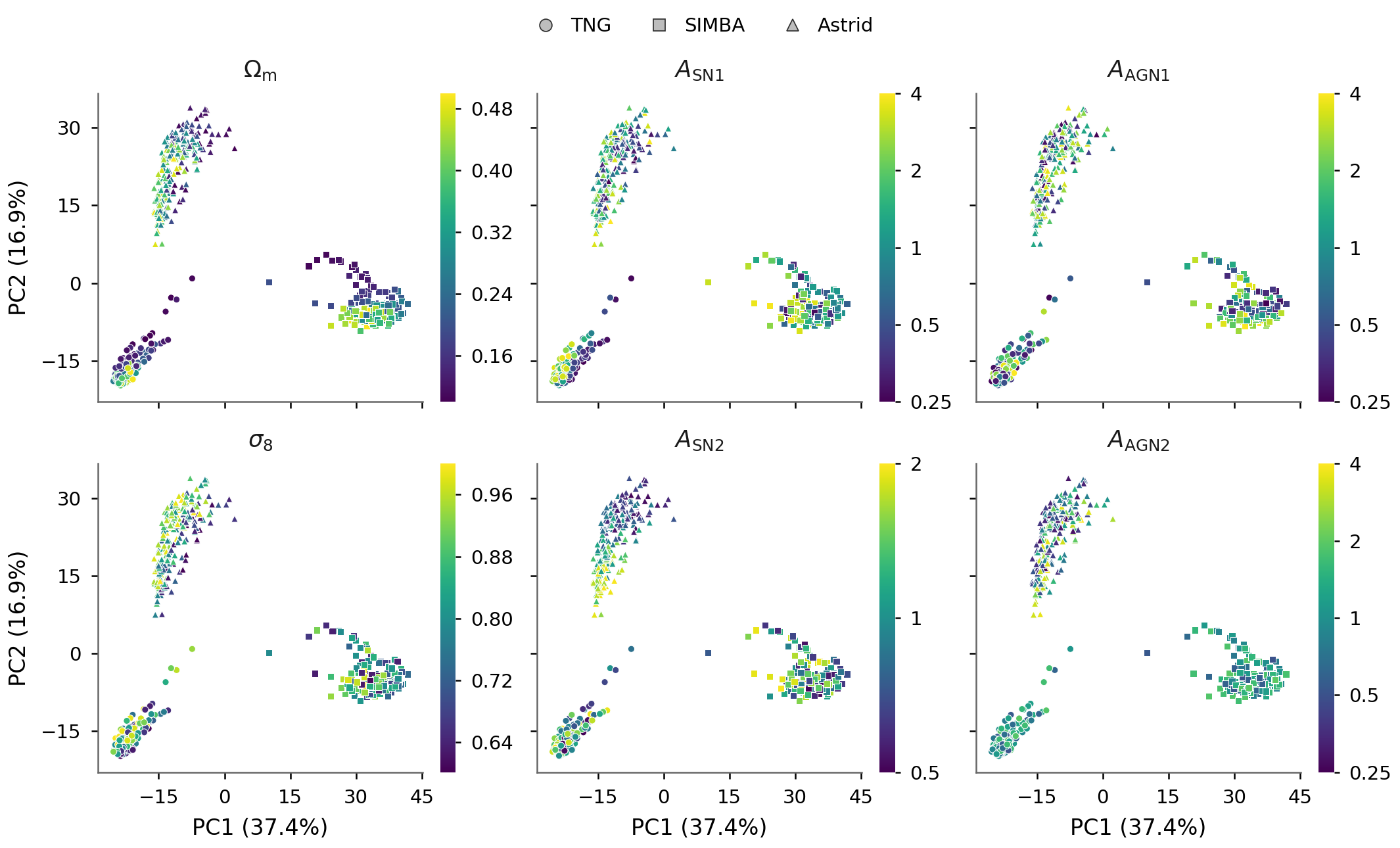}
 \caption{Catalog embeddings of the 600 test simulations (200 per suite), projected onto the first two principal components of the 2,100 training catalogs of IllustrisTNG, SIMBA and Astrid (equation~\ref{eq:catalog_pca}). Each panel shows the same catalogs colored by one simulation parameter, and marker shapes indicate the suite. The feedback parameters are shown on logarithmic color scales, and the axis labels give the fraction of variance carried by each component.}
 \label{fig:catalog_pca}
\end{figure*}

Figure~\ref{fig:catalog_pca} projects the embeddings of the 600 test catalogs onto the first two principal components of the 2,100 training catalogs of the three suites, which carry 37.4\% and 16.9\% of the total variance. The suites occupy separate regions of this plane. The first component separates SIMBA from IllustrisTNG and Astrid, and the second separates Astrid from IllustrisTNG. Note that the encoder never receives the identity of the code during training. It learns this separation automatically from the catalogs themselves, via the intrinsic differences between how the three galaxy-formation models populate their halos with galaxies, gas and black holes \citep{villaescusa2021camels,ni2023camels}.
 
Within each suite, the catalogs are ordered by $\Omega_{\mathrm m}$, and in IllustrisTNG and Astrid also by a supernova-feedback amplitude. The IllustrisTNG catalogs form an elongated sequence along which $\Omega_{\mathrm m}$ decreases from the compact lower-left end to the extended upper-right end, and $A_{\mathrm{SN1}}$ changes across the sequence. In SIMBA and Astrid, $\Omega_{\mathrm m}$ changes smoothly across the point cloud with the same trend. In Astrid, $A_{\mathrm{SN2}}$ decreases from the lower to the upper end of the cloud, and $\sigma_8$ varies across it more weakly. The AGN amplitudes show no clear trend in this plane. A linear readout of the full embedding recovers $A_{\mathrm{AGN1}}$ in SIMBA with $R^2=0.94$ and $A_{\mathrm{AGN2}}$ in Astrid with $R^2=0.996$ (Section~\ref{sec:results_readout}), so the information on AGN feedback lies along directions of lower variance.

\subsection{Linear readout of the simulation parameters}
\label{sec:results_readout}
 
\begin{figure*}[t]
 \centering
 \includegraphics[width=\textwidth]{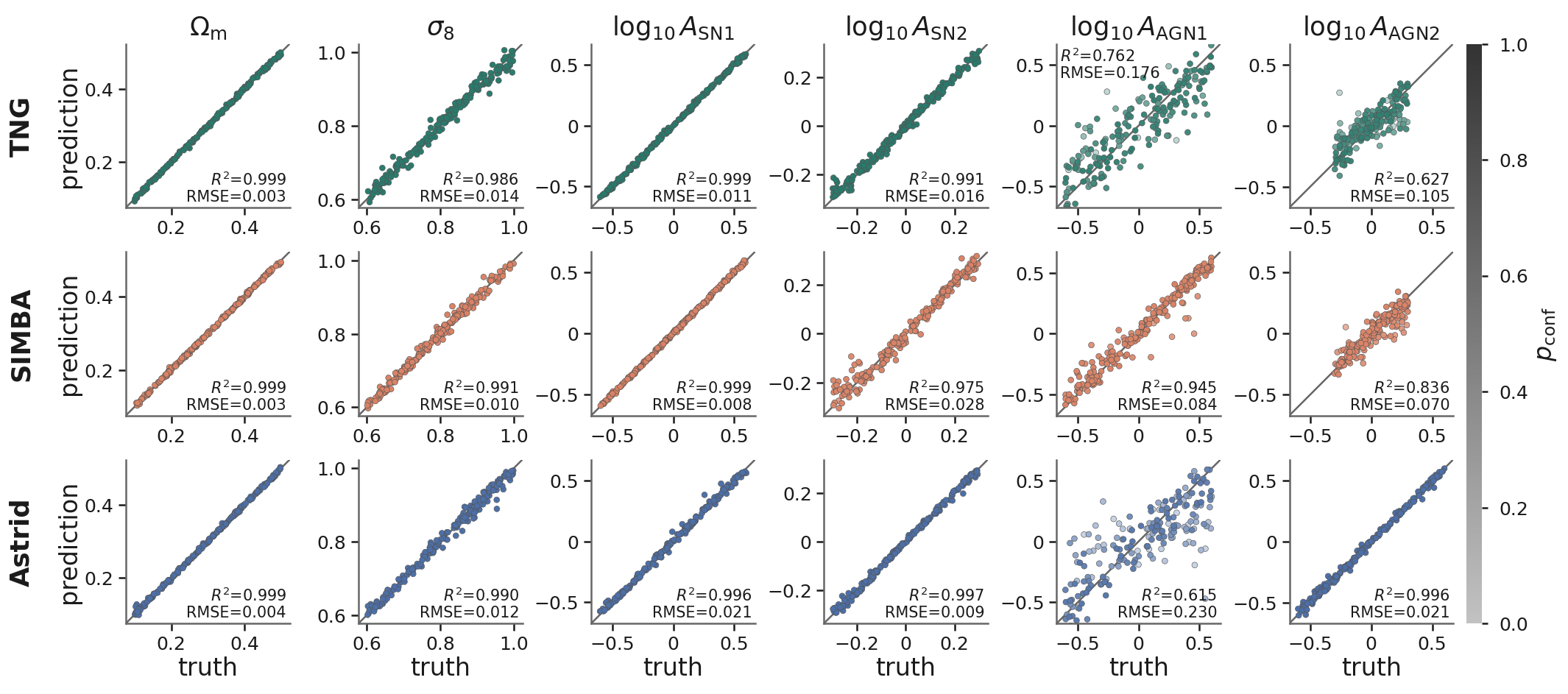}
 \caption{Linear readout of the six simulation parameters for the 200 test catalogs of each suite (rows), using all input features. Each panel compares the prediction with the true value. The feedback parameters are shown as $\log_{10}A$. The shading gives the confidence score $p_{\mathrm{conf}}$ of each prediction (equation~\ref{eq:pconf}), and the numbers give the test $R^2$ and root-mean-square error.}
 \label{fig:readout}
\end{figure*}

\begin{table}
 \centering
 \caption{Coefficient of determination $R^2$, root-mean-square error (RMSE, with the feedback parameters in $\log_{10}A$) and mean confidence score $\langle p_{\mathrm{conf}}\rangle$ of the linear readout for the 200 test catalogs of each suite.}
 \label{tab:readout}
 \setlength{\tabcolsep}{4.5pt}
 \begin{tabular}{@{}llccc@{}}
  \toprule
  Suite & Parameter & $R^2$ & RMSE & $\langle p_{\mathrm{conf}}\rangle$\\
  \midrule
  IllustrisTNG & $\Omega_{\mathrm m}$ & 0.999 & 0.003 & 0.999\\
   & $\sigma_8$ & 0.986 & 0.014 & 0.988\\
   & $A_{\mathrm{SN1}}$ & 0.999 & 0.011 & 0.999\\
   & $A_{\mathrm{SN2}}$ & 0.991 & 0.016 & 0.992\\
   & $A_{\mathrm{AGN1}}$ & 0.762 & 0.176 & 0.751\\
   & $A_{\mathrm{AGN2}}$ & 0.627 & 0.105 & 0.727\\
  \midrule
  SIMBA & $\Omega_{\mathrm m}$ & 0.999 & 0.003 & 0.999\\
   & $\sigma_8$ & 0.991 & 0.010 & 0.995\\
   & $A_{\mathrm{SN1}}$ & 0.999 & 0.008 & 0.999\\
   & $A_{\mathrm{SN2}}$ & 0.975 & 0.028 & 0.967\\
   & $A_{\mathrm{AGN1}}$ & 0.945 & 0.084 & 0.953\\
   & $A_{\mathrm{AGN2}}$ & 0.836 & 0.070 & 0.909\\
  \midrule
  Astrid & $\Omega_{\mathrm m}$ & 0.999 & 0.004 & 0.999\\
   & $\sigma_8$ & 0.990 & 0.012 & 0.992\\
   & $A_{\mathrm{SN1}}$ & 0.996 & 0.021 & 0.999\\
   & $A_{\mathrm{SN2}}$ & 0.997 & 0.009 & 0.999\\
   & $A_{\mathrm{AGN1}}$ & 0.615 & 0.230 & 0.591\\
   & $A_{\mathrm{AGN2}}$ & 0.996 & 0.021 & 0.997\\
  \bottomrule
 \end{tabular}
\end{table}

For each suite and parameter, we fit a ridge regression on the embeddings of the 700 training catalogs, and apply it to the 200 test catalogs. Figure~\ref{fig:readout} compares the predictions with the true values, and Table~\ref{tab:readout} lists the test $R^2$ and root-mean-square error (RMSE). A linear map recovers the cosmological and supernova-feedback parameters with $R^2\geq0.97$ in all three suites, and the AGN parameters with a precision that depends on the code.

\paragraph{Cosmological parameters.} The matter density is recovered almost exactly, with $R^2=0.999$ in all three suites and an RMSE of 0.003 to 0.004, below 1\% of the prior range $0.1\leq\Omega_{\mathrm m}\leq0.5$. This $\Omega_{\mathrm m}$ information is likely carried by two components: the standard large-scale clustering signal in the galaxy positions \citep{hahn2023simbig} as well as the $\Omega_{\mathrm m}$-modulated correlations in individual galaxy properties. The latter have previously been shown empirically in the CAMELS simulations \citep{villaescusa2022onegalaxy, hahn2024cosmology} and explained via the mechanism that galaxy properties constrain the universal baryon fraction $\Omega_{\mathrm b}/\Omega_{\mathrm m}$, which, when $\Omega_{\mathrm b}$ is held constant in CAMELS, provides precise constraints on $\Omega_{\mathrm m}$ \citep{lue2025cosmology, liao2026onegalaxy}. For $\sigma_8$, $R^2=0.986$--$0.991$ and the RMSE is $0.010$--$0.014$. The readouts reach nearly the same precision in the three codes, which form galaxies differently. A readout fitted on Swift-EAGLE labels reaches the same precision in a code that the encoder never saw (Section~\ref{sec:results_swift}). The $\sigma_8$ constraints are similar to those of $\Omega_{\mathrm m}$, combining information from large-scale clustering as well as ensemble statistics of galaxy populations, driven by differences in merger histories and environments \citep{chawak2024multiple, lovell2025luminosity}.

\paragraph{Supernova feedback.} The two supernova-feedback amplitudes are recovered with $R^2\geq0.97$ in every suite, $A_{\mathrm{SN1}}$ with $R^2=0.996$--$0.999$ and $A_{\mathrm{SN2}}$ with $0.975$--$0.997$. Galactic winds regulate the stellar mass, gas content and metallicity of nearly every galaxy, so their strength leaves an imprint on most of the objects that the encoder sees. In the one-parameter variations, raising a supernova parameter to its largest value also moves the catalogs furthest in every suite. (Section~\ref{sec:results_response}).
 
\paragraph{AGN feedback.} The precision for the AGN amplitudes depends on the code and on the feedback mode that they scale. In SIMBA, $A_{\mathrm{AGN1}}$ and $A_{\mathrm{AGN2}}$ reach $R^2=0.94$ and $0.84$, and in Astrid the thermal-mode amplitude $A_{\mathrm{AGN2}}$ reaches $0.996$. The three parameters with the lowest $R^2$ are $A_{\mathrm{AGN1}}$ and $A_{\mathrm{AGN2}}$ in IllustrisTNG ($0.76$ and $0.63$) and $A_{\mathrm{AGN1}}$ in Astrid ($0.62$), and all three scale the kinetic feedback mode. This mode operates mainly in massive black holes. In IllustrisTNG, the Eddington ratio below which a black hole switches to kinetic feedback grows with black-hole mass about a pivot of $10^8\,\mathrm M_\odot$ \citep{weinberger2017agn}, and black holes in Astrid enter the kinetic mode above $5\times10^8\,h^{-1}\,\mathrm M_\odot$ \citep{ni2023camels}. Only a small fraction of the subhalos in a $25\,h^{-1}\,\mathrm{Mpc}$ box host such black holes, so these amplitudes act directly on few of the objects in a catalog. Section~\ref{sec:results_pconf} shows that the information on them grows with the number of black holes above these masses.

\paragraph{Confidence score.} Each prediction carries an error bar and a confidence score $p_{\mathrm{conf}}$, both computed from the embedding alone (Section~\ref{sec:pconf}). The score is one minus the squared ratio of the half-width of the 68\% error bar to the standard deviation of the parameter over the training catalogs (equation~\ref{eq:pconf}). Like $R^2$, it approaches 1 when the error bar is small compared with the spread of the parameter and 0 when it is as wide as that spread.

The shading in Fig.~\ref{fig:readout} shows $p_{\mathrm{conf}}$ for every test catalog, and Table~\ref{tab:readout} lists its mean. It is close to 1 for the cosmological and supernova parameters and for $A_{\mathrm{AGN2}}$ in Astrid. For the other AGN parameters, it varies from catalog to catalog, and the catalogs with low $p_{\mathrm{conf}}$ scatter most widely about the diagonal. In SIMBA, the five catalogs with the largest errors in each AGN parameter all lie in the quarter of catalogs with the lowest $p_{\mathrm{conf}}$. Section~\ref{sec:results_pconf} relates $p_{\mathrm{conf}}$ to the accuracy of the readout and to the black-hole population of each catalog, and Appendix~\ref{app:pit} tests the calibration of the error bars.

\subsection{Restricted inputs and comparison with supervised networks}
\label{sec:results_restricted}
 
\begin{figure*}[t]
 \centering
 \includegraphics[width=\textwidth]{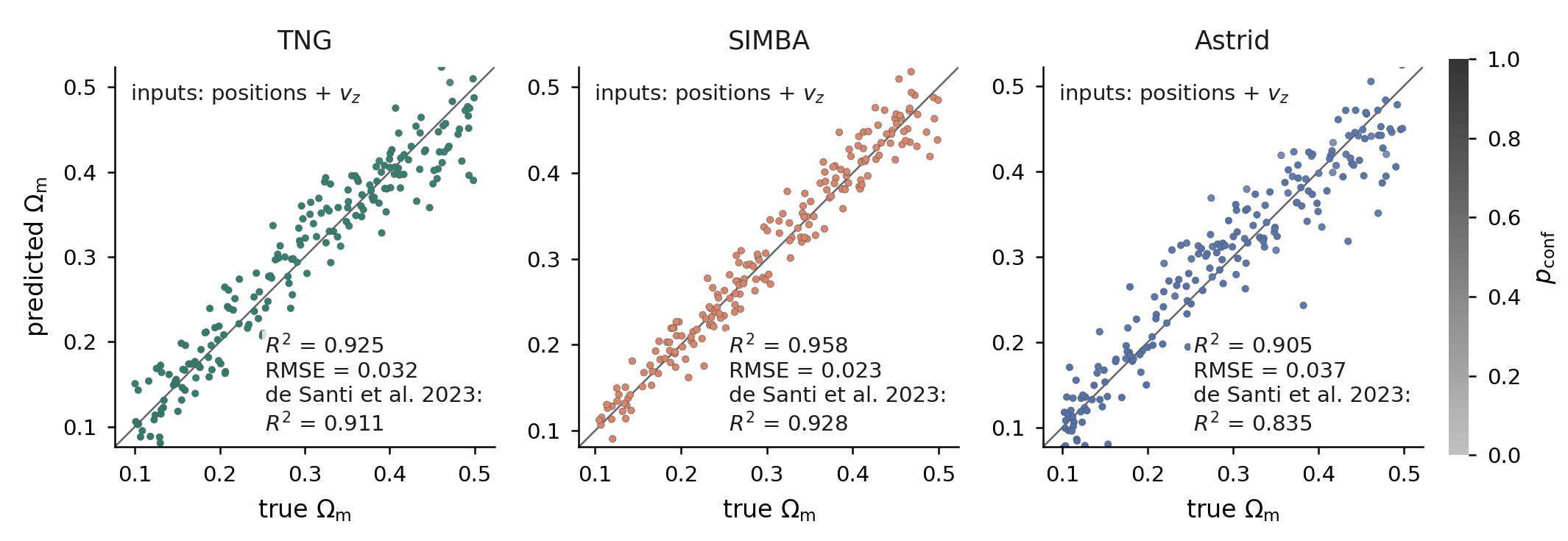}
 \caption{$\Omega_{\mathrm m}$ from the 3D positions and line-of-sight velocities $v_z$ of galaxies with $M_\star>1.95\times10^8\,h^{-1}\,\mathrm M_\odot$ for the 200 test catalogs of each suite. The frozen encoder sees only these features, and a linear readout is fitted on the training catalogs. The quoted values from \citet{desanti2023robust} are for their GNNs, trained and tested on their own split of the same suite. The shading gives $p_{\mathrm{conf}}$.}
 \label{fig:desanti}
\end{figure*}
 
\begin{figure*}[t]
 \centering
 \includegraphics[width=\textwidth]{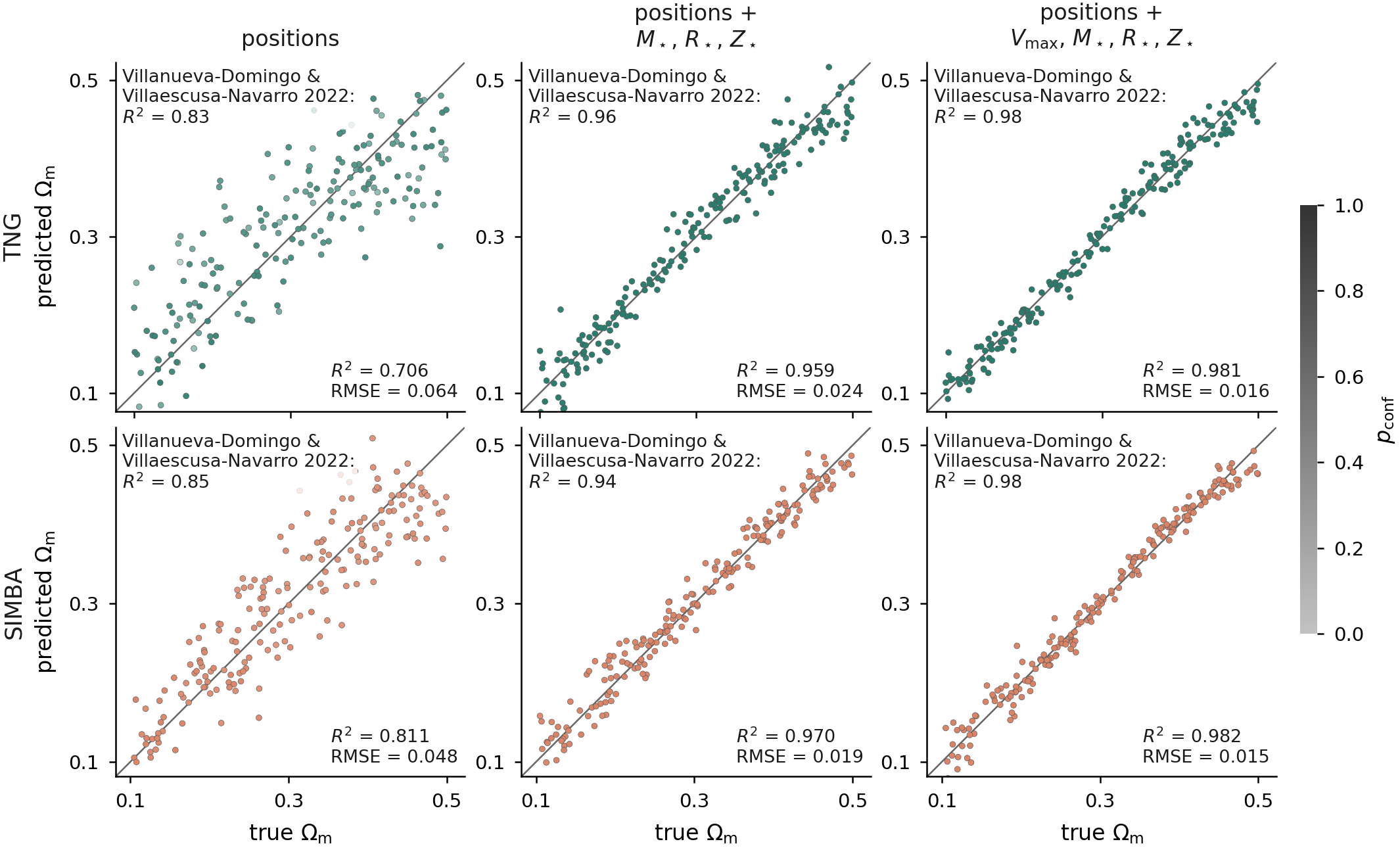}
 \caption{$\Omega_{\mathrm m}$ in IllustrisTNG (top) and SIMBA (bottom) from the subhalos with at least 20 star particles, using positions alone (left), positions with stellar mass, stellar half-mass radius and stellar metallicity (middle), and these with $V_{\max}$ added (right). The frozen encoder sees only these features, and a linear readout is fitted on the training catalogs. The numbers in the lower right give the test $R^2$ and root-mean-square error of the readout, and those in the upper left the $R^2$ of the GNNs of \citet{villanueva2022cosmicgraphs} with the same inputs, on their own test catalogs. The shading gives $p_{\mathrm{conf}}$.}
 \label{fig:vd22}
\end{figure*}

The frozen encoder also embeds catalogs in which only some features are available, because it was trained to represent catalogs with randomly hidden inputs. We use this property to compare the embedding with the supervised graph neural networks (GNNs) of two earlier studies, which infer $\Omega_{\mathrm m}$ from restricted galaxy catalogs. For each input set, we select galaxies by their stellar content as described below, without the dark-matter threshold of Section~\ref{sec:data_catalogs}, hide all other features, embed the training and test catalogs with the unchanged encoder, and fit a new linear readout on the training catalogs.

\citet{desanti2023robust} trained GNNs on the 3D positions and line-of-sight velocities of galaxies with $M_\star>1.95\times10^8\,h^{-1}\,\mathrm M_\odot$ and tested them on catalogs of the same suite. With the same input, the linear readout of our embedding reaches $R^2=0.925$, $0.958$ and $0.905$ in IllustrisTNG, SIMBA and Astrid (Fig.~\ref{fig:desanti}), compared with $0.911$, $0.928$ and $0.835$ for their GNNs.
 
\citet{villanueva2022cosmicgraphs} used positions alone, positions with stellar mass, stellar half-mass radius and stellar metallicity, and these with $V_{\max}$ added, in IllustrisTNG and SIMBA. We give the encoder the same three input sets for all subhalos with at least 20 star particles (Fig.~\ref{fig:vd22}). With positions alone, their GNNs reach $R^2=0.83$ and $0.85$ and our readout $0.71$ and $0.81$. With the stellar properties, the embedding reaches $R^2=0.96$ and $0.97$, comparable to the $0.96$ and $0.94$ of their GNNs, and with $V_{\max}$ added the embedding and the GNNs both reach $R^2=0.98$ in each suite.

Once the positions are complemented by the line-of-sight velocity (Fig.~\ref{fig:desanti}) or by the stellar properties (Fig.~\ref{fig:vd22}), the linear readout of a single encoder trained without labels reaches a precision comparable to that of the GNNs, which were each trained for one parameter and one input set. Section~\ref{sec:disc_objective} discusses the case of positions alone. The precision grows with every property that becomes visible. In IllustrisTNG, the RMSE of $\Omega_{\mathrm m}$ falls from 0.064 with positions alone to 0.024 with the stellar properties and 0.016 with $V_{\max}$ added, and in SIMBA from 0.048 to 0.019 and 0.015. With all input features of every subhalo, the full catalogs give 0.003 in both suites, and the same embedding carries information on all six parameters.

\subsection{Transfer to an unseen simulation code}
\label{sec:results_swift}
 
\begin{figure*}[t]
 \centering
 \includegraphics[width=\textwidth]{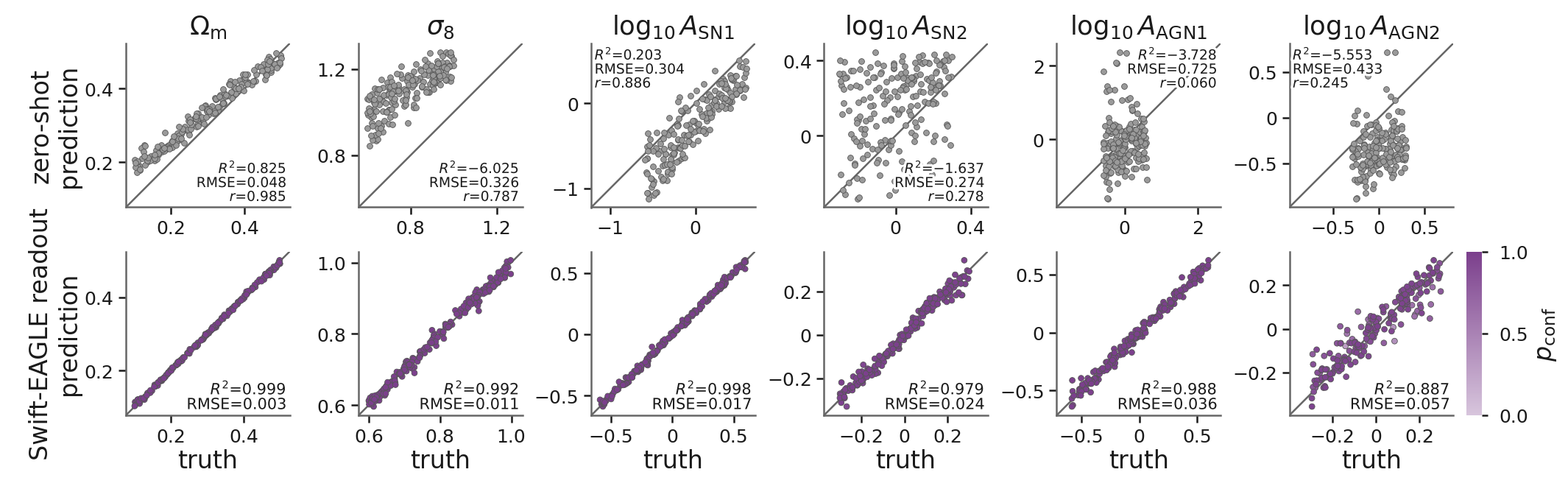}
 \caption{Readout of the 200 Swift-EAGLE test catalogs. The encoder never saw this code during training. Top: zero-shot predictions of the readout fitted on the training catalogs of IllustrisTNG, SIMBA and Astrid. Bottom: predictions of a readout fitted on the 700 Swift-EAGLE training catalogs, with the shading giving $p_{\mathrm{conf}}$. The numbers give the test $R^2$, the root-mean-square error and, for the zero-shot predictions, the Pearson correlation $r$.}
 \label{fig:swift}
\end{figure*}
 
\begin{figure*}[t]
 \centering
 \includegraphics[width=\textwidth]{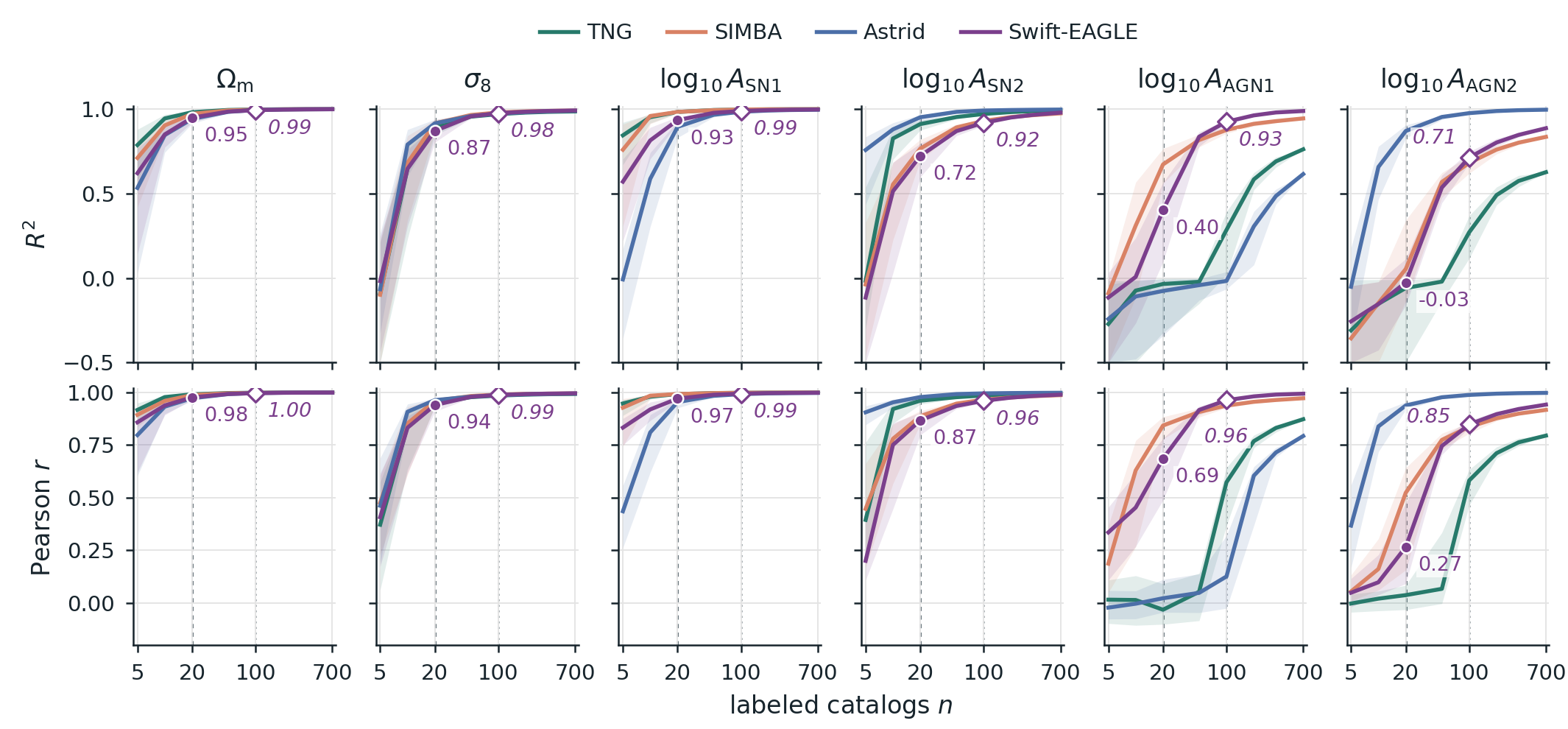}
 \caption{Test $R^2$ (top) and Pearson correlation $r$ (bottom) as a function of the number $n$ of labeled training catalogs used to fit the readout, for the three training suites and Swift-EAGLE. Lines show the median over 100 random draws of the $n$ catalogs and bands the 16th to 84th percentiles, and $R^2$ below $-0.5$ is shown at $-0.5$. Markers give the Swift-EAGLE values at $n=20$ and 100.}
 \label{fig:label_efficiency}
\end{figure*}

The Swift-EAGLE suite combines the \textsc{swift} code with a galaxy-formation model based on EAGLE (Section~\ref{sec:data_simulations}), and its catalogs were used in neither the self-supervised training nor the input normalization. We embed them with the frozen encoder and read them out in two ways (Fig.~\ref{fig:swift}).
 
\paragraph{Zero-shot readout.} We apply the pooled readout of the training suites to the Swift-EAGLE catalogs. The $\Omega_{\mathrm m}$ predictions correlate with the true values with $r=0.985$. They follow a line of slope 0.77 and lie above the true values by 0.04 on average, which gives $R^2=0.825$. The $\sigma_8$ predictions lie above the true values by 0.32 on average, with $r=0.79$. The $A_{\mathrm{SN1}}$ predictions follow the true values with a slope of 1.00 and $r=0.89$ and lie below them by 0.25 in $\log_{10}A_{\mathrm{SN1}}$, which gives $R^2=0.20$. The embedding of the new code thus varies with $\Omega_{\mathrm m}$, $\sigma_8$ and $A_{\mathrm{SN1}}$ along directions that the readout of the training suites recognizes, and each of these predictions needs a new calibration. The training suites share their response directions for the same three parameters (Section~\ref{sec:results_response}). The other feedback parameters do not transfer ($r\leq0.28$). In Swift-EAGLE, $A_{\mathrm{SN2}}$, $A_{\mathrm{AGN1}}$ and $A_{\mathrm{AGN2}}$ scale the metallicity dependence of the stellar feedback energy, the Bondi accretion rate and the heating temperature of AGN feedback, and in the training codes the same amplitudes scale other processes (Table~\ref{tab:data_suites}).

\paragraph{Readout with Swift-EAGLE labels.} With labels, the embedding of Swift-EAGLE is as informative as those of the training suites. A linear readout fitted on its 700 training catalogs reaches $R^2=0.999$, 0.992, 0.998 and 0.979 for $\Omega_{\mathrm m}$, $\sigma_8$, $A_{\mathrm{SN1}}$ and $A_{\mathrm{SN2}}$ on its 200 test catalogs. For $A_{\mathrm{AGN1}}$ and $A_{\mathrm{AGN2}}$, it reaches 0.988 and 0.887, higher than in IllustrisTNG and SIMBA.
 
\paragraph{Label efficiency.} Figure~\ref{fig:label_efficiency} shows how many labeled catalogs these readouts need. For each $n<700$, we draw 100 random sets of $n$ training catalogs, choose $\alpha$ and fit the readout on each set, and evaluate each fit on the test catalogs. Besides $R^2$, we show the Pearson correlation $r$ between predicted and true values, which measures how well the predictions follow the true values up to a linear calibration. For Swift-EAGLE, 20 labeled catalogs give $r\geq0.94$ for $\Omega_{\mathrm m}$, $\sigma_8$ and $A_{\mathrm{SN1}}$, and 100 give $r\geq0.96$ for all parameters except $A_{\mathrm{AGN2}}$ ($r=0.85$). With 20 labels, $\Omega_{\mathrm m}$ reaches $R^2\geq0.94$ and $\sigma_8$ reaches $R^2\geq0.87$ in all four codes. Of the cosmological and supernova parameters, $A_{\mathrm{SN2}}$ needs the most additional labels in Swift-EAGLE, with $R^2=0.72$ for 20 labels against 0.77 to 0.95 in the training suites. The Swift-EAGLE curves fall within or close to the range of the training suites, so the new code needs about as many labels as the codes on which the encoder was trained.

The number of labels depends on how strongly a parameter imprints on the catalogs. With 100 labels, $A_{\mathrm{AGN1}}$ reaches $R^2=0.93$ in Swift-EAGLE, 0.88 in SIMBA, 0.28 in IllustrisTNG and about zero in Astrid. For the three kinetic-mode parameters, the readout finds no signal with 50 or fewer labeled catalogs ($|r|\leq0.07$). Their $r$ rises steeply between 50 and 200 catalogs, and their $R^2$ is still increasing at $n=700$.
 
For a new code, the cosmological directions of the embedding carry over without labels, and about 100 labeled catalogs recover its feedback parameters with $r\gtrsim0.85$.

%% file: 04-Results_02.tex
\subsection{Response to parameters and to initial conditions}
\label{sec:results_response}
 
\begin{figure*}[t]
 \centering
 \includegraphics[width=\textwidth]{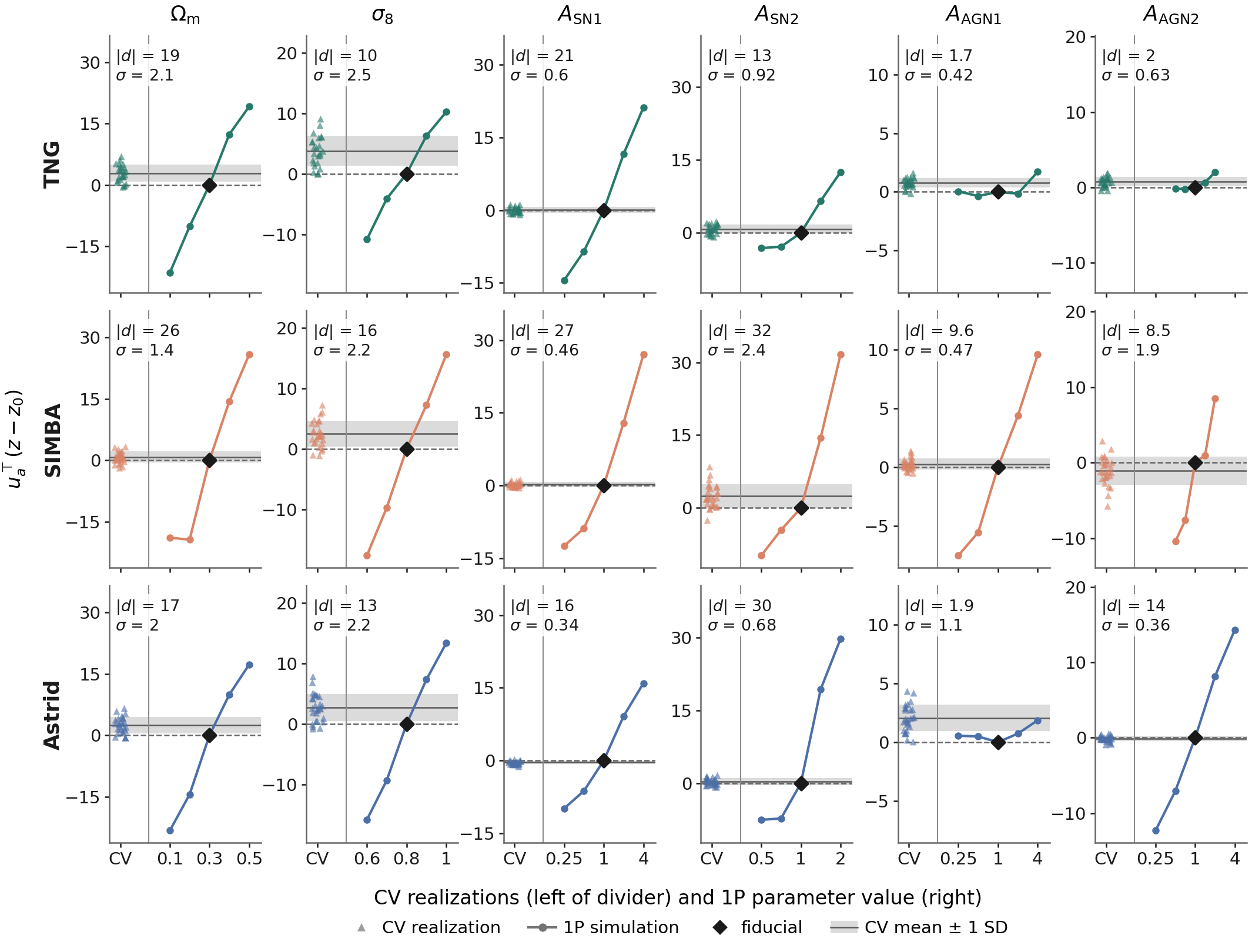}
 \caption{Response of the catalog embedding to each simulation parameter (columns) in each suite (rows). For parameter $a$, the vertical axis is the projection $\boldsymbol u_a^{\mathsf T}(\boldsymbol z-\boldsymbol z_0)$ onto the direction from the fiducial 1P catalog (diamond) to the 1P catalog with the largest value of $a$. Circles show the 1P catalogs and triangles the CV catalogs (Section~\ref{sec:data_simulations}), with the mean and standard deviation of the CV catalogs as line and gray band. Each panel gives the length $|\boldsymbol d|=\|\boldsymbol d_a\|$ of the displacement and the standard deviation $\sigma=\sigma_a$ of the CV catalogs along $\boldsymbol u_a$.}
 \label{fig:response}
\end{figure*}
 
\begin{figure*}[t]
 \centering
 \includegraphics[width=\textwidth]{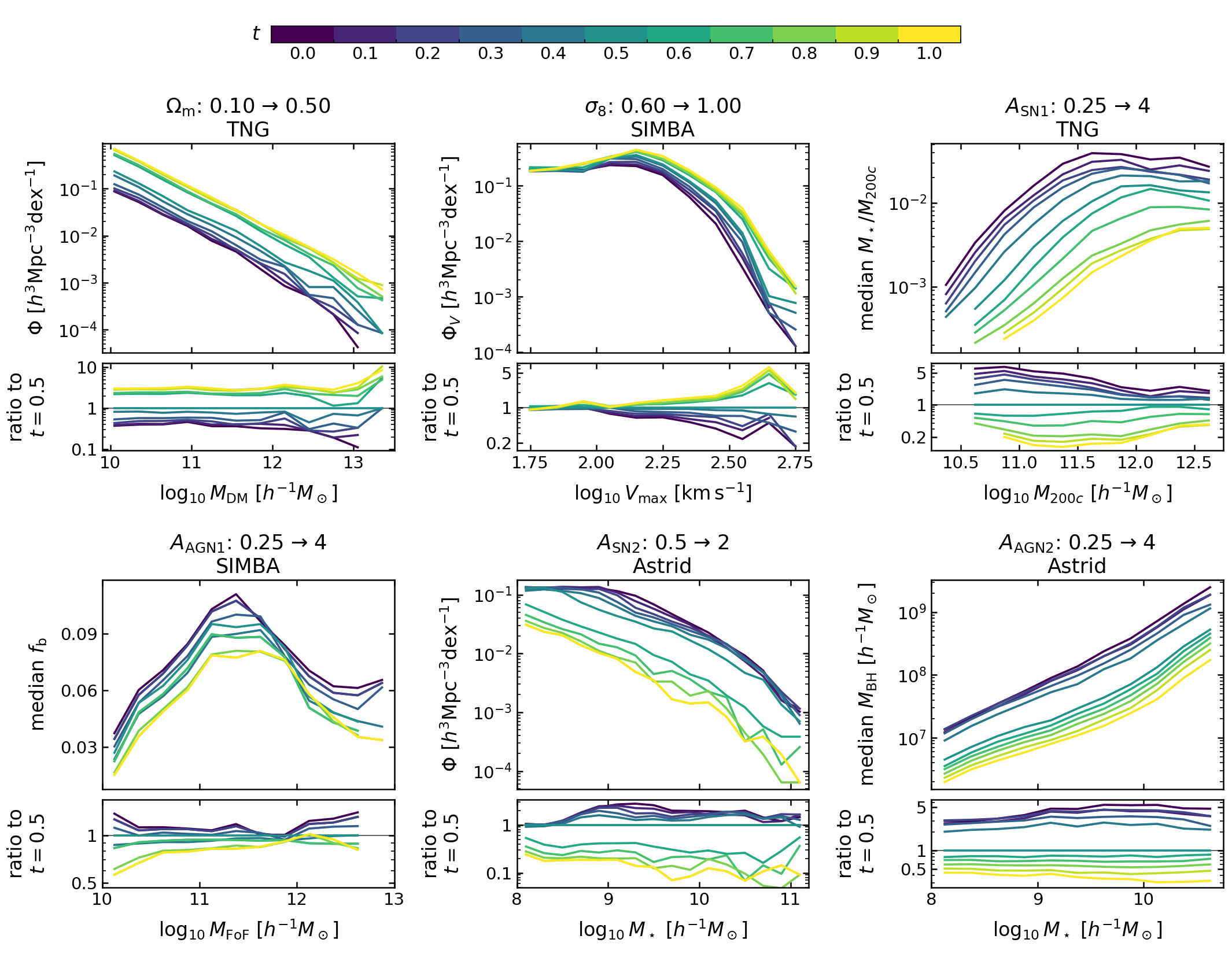}
 \caption{Statistics of the LH catalogs retrieved along straight lines in the embedding. Each panel follows one parameter in one suite (title). The line runs from the embedding of the 1P catalog with the lowest value of the parameter ($t=0$) to that of the 1P catalog with the highest value ($t=1$), and each of 11 equally spaced points on it is represented by the mean statistic of the five nearest of all 1,000 LH catalogs of the same suite. $\Phi$ is the number of subhalos per unit volume and per dex in dark-matter mass, $V_{\max}$ or stellar mass. The other panels show medians in bins of halo or stellar mass, with the stellar-to-halo mass ratio taken for central galaxies and $f_{\mathrm b}$ the baryon fraction of a FoF halo. Only bins populated in at least half of the training catalogs are shown. The lower part of each panel gives the ratio to the midpoint of the line, $t=0.5$.}
 \label{fig:interp}
\end{figure*}
 
\begin{figure*}[t]
 \centering
 \includegraphics[width=0.8\textwidth]{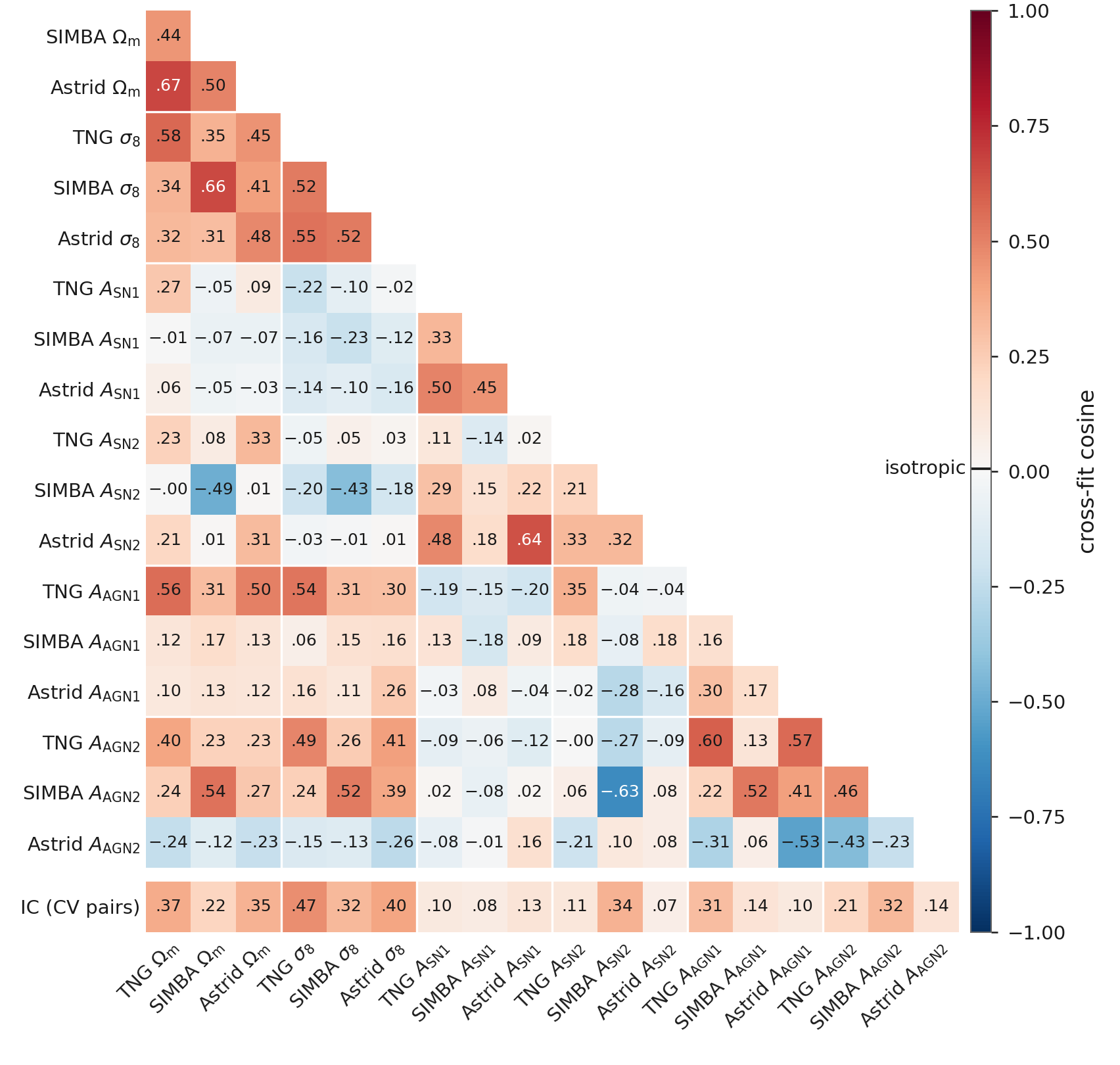}
 \caption{Cross-fitted cosine similarity between the response directions of the six parameters in the three suites, estimated from all 1,000 LH catalogs of each suite. Red cells mark parameters that move catalogs in similar directions and blue cells opposite directions, and isotropic random directions give zero. The bottom row (IC, initial conditions) gives the root-mean-square cosine between each direction and the displacements between pairs of CV catalogs of the same suite. The mark on the color bar shows its value for isotropic random directions.}
 \label{fig:similarity}
\end{figure*}

The geometry of the embedding reflects the physical effect of each parameter. The 1P and CV sets of CAMELS  probe this geometry around the fiducial model. Here we compute $\boldsymbol z$ from equation~\ref{eq:catalog_pca} with $\mathcal T$ set to the 2,100 training catalogs of the three suites and $s_{\mathcal T}=1$, and keep its 2,048 leading components. For each parameter $a$, we take the displacement $\boldsymbol d_a=\boldsymbol z(\theta_a^{\max})-\boldsymbol z_0$ from the fiducial 1P catalog to the one with the largest value of $a$, and project the 1P and CV catalogs onto $\boldsymbol u_a=\boldsymbol d_a/\|\boldsymbol d_a\|$ (Fig.~\ref{fig:response}). The length $\|\boldsymbol d_a\|$ measures the response distance, and the standard deviation $\sigma_a$ of the CV catalogs along $\boldsymbol u_a$ measures the scatter from initial conditions.

\paragraph{Response and cosmic variance.} Apart from the three kinetic-mode amplitudes discussed below, every parameter moves a catalog well beyond this scatter, and its five 1P catalogs follow the parameter value monotonically along $\boldsymbol u_a$, with a single exception at the two lowest values of $\Omega_{\mathrm m}$ in SIMBA. The supernova parameters shift the catalog by 13 to 59 times $\sigma_a$, $\Omega_{\mathrm m}$ by 9 to 19 times and $\sigma_8$ by 4 to 7 times. The AGN parameters do so in SIMBA (20 and 4.6 times), and $A_{\mathrm{AGN2}}$ does so in Astrid (39 times). For $A_{\mathrm{AGN1}}$ and $A_{\mathrm{AGN2}}$ in IllustrisTNG and for $A_{\mathrm{AGN1}}$ in Astrid, the four lower values leave the catalog within about $\sigma_a$ of the fiducial catalog, and only the largest value moves it, by $\|\boldsymbol d_a\|=1.7$, 2.0 and 1.9, or 4.0, 3.2 and 1.6 times $\sigma_a$. These displacements are at least four times shorter than those of all other parameters. The 1P response alone thus singles out the three parameters with the lowest readout $R^2$.

\paragraph{Physical content of the response.} Figure~\ref{fig:interp} shows how the catalogs change along these displacements. For one parameter in one suite per panel, we interpolate 11 equally spaced points on the straight line in the embedding from the 1P catalog with the lowest value of the parameter to the one with the highest value, and represent each point by the five nearest of all 1,000 LH catalogs of the same suite. Because the retrieved catalogs are simulations, their statistics show what each point of the line represents. Along every line, the statistics shift from one end to the other in the direction expected for the parameter.

Increasing $\Omega_{\mathrm m}$ from 0.1 to 0.5 raises the abundance of subhalos in IllustrisTNG with dark-matter masses of $10^{11}$--$10^{12}\,h^{-1}\,\mathrm M_\odot$ by factors of 8 to 9. Increasing $\sigma_8$ from 0.6 to 1.0 raises the abundance of SIMBA subhalos by a factor of 1.4 at $V_{\max}\approx110\,\mathrm{km\,s^{-1}}$ and of 12 at $350\,\mathrm{km\,s^{-1}}$. The wind energy $A_{\mathrm{SN1}}$ lowers the median ratio of stellar to halo mass of IllustrisTNG central galaxies by a factor of about 40 at $M_{200\mathrm c}=10^{11}$--$10^{11.5}\,h^{-1}\,\mathrm M_\odot$ and of 9 at $10^{12}$--$10^{12.25}\,h^{-1}\,\mathrm M_\odot$, and the wind speed $A_{\mathrm{SN2}}$ lowers the abundance of Astrid galaxies with $M_\star=10^{9.2}$--$10^{9.8}\,h^{-1}\,\mathrm M_\odot$ by factors of 20 to 28. The AGN parameters act on the gas and on the black holes. $A_{\mathrm{AGN1}}$ lowers the median baryon fraction of SIMBA FoF halos at all masses shown, by 10\% to 60\% and least near $M_{\mathrm{FoF}}=10^{12}\,h^{-1}\,\mathrm M_\odot$. The thermal-mode amplitude $A_{\mathrm{AGN2}}$ lowers the median black-hole mass of Astrid galaxies at fixed stellar mass by a factor of about 7 at $M_\star=10^{8}$--$10^{8.5}\,h^{-1}\,\mathrm M_\odot$ and by factors of 14 to 18 at $10^{9.75}$--$10^{10.75}\,h^{-1}\,\mathrm M_\odot$. We choose these statistics for their strong response to the parameters, and the same retrieval applies to any statistic that the catalogs provide. A systematic study of how observables respond to each parameter is left for future work.

\paragraph{Directions of the response.} Figure~\ref{fig:similarity} compares the directions in which the parameters move the catalogs across the Latin hypercube. For each suite, we regress the covariance vectors $\boldsymbol\xi$ of all 1,000 LH catalogs jointly on the standardized targets of equation~(\ref{eq:readout_targets}), and the coefficient vector of each parameter is its response direction. Noise in the estimated directions biases their norms upwards, and it also biases the inner products between directions of the same suite. We therefore fit every direction on 200 random splits of the catalogs of its suite into two halves, and take the norms and the within-suite inner products from the mean product of the two half-sample estimates.

The cosmological directions are shared across codes. The $\Omega_{\mathrm m}$ directions of the three suites have cosines of 0.44 to 0.67 with one another and the $\sigma_8$ directions 0.52 to 0.55, and within each code the two are partly aligned (0.48 to 0.66). This is expected, because gravity and initial conditions act in the same way in all codes and both parameters raise the abundance of massive halos. For the astrophysical parameters, we compare the directions with the responses of galaxies, black holes and gas in the 1P simulations.

Within a code, the alignments agree with these responses. In SIMBA, the momentum flux $A_{\mathrm{AGN1}}$ and the jet speed $A_{\mathrm{AGN2}}$ align (0.52), and the wind speed $A_{\mathrm{SN2}}$ points against the jet speed ($-0.63$). Both AGN parameters spread the gas of the SIMBA 1P set farther from the dark matter next to which it started, and faster winds reduce this spread \citep[Fig.~4 of][]{gebhardt2024spread}, because stellar feedback limits the growth of black holes and thereby the output of the jets \citep{delgado2023clustering,gebhardt2024spread}. In IllustrisTNG, the energy $A_{\mathrm{AGN1}}$ and the ejection speed $A_{\mathrm{AGN2}}$ of the kinetic mode align (0.60), and both lower the baryon fractions of halos \citep[Fig.~2 of][]{delgado2023clustering}. In Astrid, the wind energy $A_{\mathrm{SN1}}$ and the wind speed $A_{\mathrm{SN2}}$ align (0.64), and both reduce the abundance of galaxies \citep[Fig.~5 of][]{ni2023camels}.

Across codes, parameters that scale similar processes align. The $A_{\mathrm{SN1}}$ directions have cosines of 0.33 to 0.50, and a larger $A_{\mathrm{SN1}}$ reduces the abundance of galaxies at all masses in all three codes \citep[Fig.~5 of][]{ni2023camels}. The largest of these cosines joins IllustrisTNG and Astrid, where $A_{\mathrm{SN1}}$ scales the same wind energy per unit star-formation rate. The ejection speed of the kinetic mode in IllustrisTNG aligns with the jet speed of SIMBA (0.46), and both lower the baryon fractions of massive halos \citep[Fig.~2 of][]{delgado2023clustering}. The kinetic-mode energy $A_{\mathrm{AGN1}}$ of Astrid aligns with the same ejection speed (0.57). Astrid and IllustrisTNG both inject kinetic feedback locally and without collimation around massive black holes at low accretion rates \citep{ni2023camels}, and in both codes the confidence of the readout follows the number of black holes massive enough to drive this mode (Section~\ref{sec:results_pconf}).

The thermal-mode amplitude $A_{\mathrm{AGN2}}$ of Astrid points against the kinetic directions of Astrid and IllustrisTNG ($-0.53$, $-0.31$ and $-0.43$) and against the jet speed of SIMBA ($-0.23$). A larger thermal-mode amplitude limits the formation of the massive black holes that turn on the kinetic mode and raises the star-formation rate and the abundance of galaxies in Astrid \citep[Figs.~4 and 5 of][]{ni2023camels}. This is expected from self-regulated black-hole growth, in which more efficient feedback leaves black holes less massive \citep{booth2009agn}. Section~\ref{sec:results_axes} follows this effect to the numbers of massive galaxies and black holes.
 
With responses of 47 and 44 times the scatter from initial conditions, the aligned supernova parameters of Astrid are both recovered with $R^2>0.99$. The kinetic-mode parameters of IllustrisTNG move the catalog by about a tenth of the displacement for $\Omega_{\mathrm m}$. The direction of $A_{\mathrm{AGN1}}$ also overlaps with those of $\Omega_{\mathrm m}$ and $\sigma_8$ (0.56 and 0.54) and with the displacements between CV catalogs (0.31), so changes of cosmology or of the initial conditions can mimic it.

\paragraph{Initial conditions.} The bottom row of Fig.~\ref{fig:similarity} compares each response direction with the displacements between CV catalogs, which are the directions along which a change of initial conditions moves a catalog. The root-mean-square cosine is 0.32 to 0.47 for $\sigma_8$ and 0.22 to 0.37 for $\Omega_{\mathrm m}$, and it is at most 0.14 for eight of the twelve feedback directions. In a $25\,h^{-1}\,\mathrm{Mpc}$ volume, the number of rare massive halos varies from one realization to the next, and this cosmic variance mimics a change of cosmology. Among the feedback parameters, $A_{\mathrm{SN2}}$ and $A_{\mathrm{AGN2}}$ in SIMBA and $A_{\mathrm{AGN1}}$ in IllustrisTNG overlap with the initial conditions about as much as $\sigma_8$ does in SIMBA (0.34, 0.32 and 0.31). The response to $A_{\mathrm{SN1}}$ points in directions that cosmic variance hardly reaches (at most 0.13).
 
The precision of a readout thus follows from the length of the response to a parameter and from the overlap of its direction with those of the other parameters and of the initial conditions.

\subsection{Confidence score and black-hole content}
\label{sec:results_pconf}
 
\begin{figure*}[t]
 \centering
 \includegraphics[width=\textwidth]{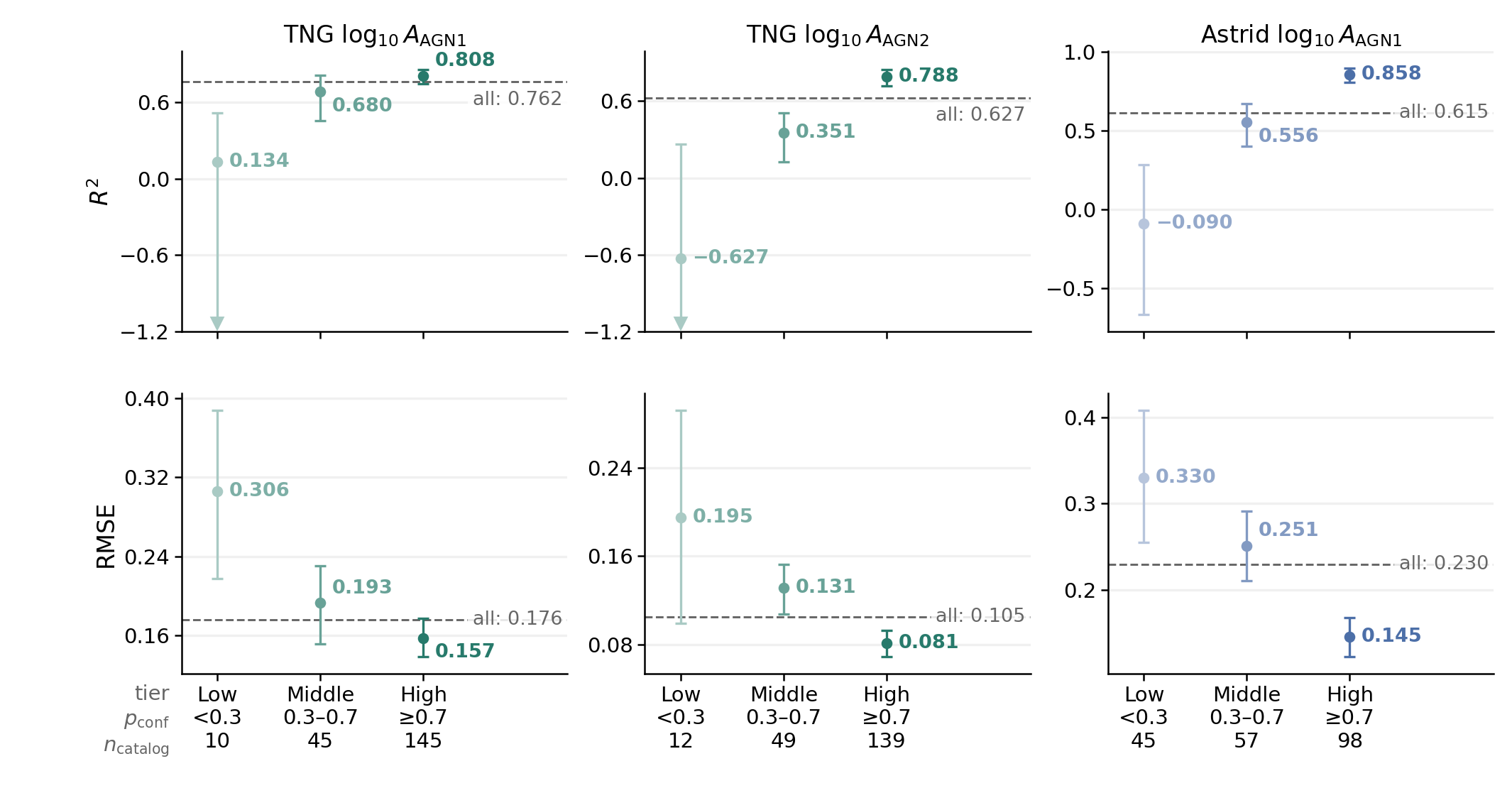}
 \caption{Test $R^2$ (top) and root-mean-square error (bottom) in each confidence tier for the three parameters with the lowest overall $R^2$. $R^2$ is computed with the variance of the true values within the tier. Dashed lines give the values for all 200 test catalogs. Error bars are 95\% bootstrap intervals from 5000 resamples of the catalogs in each tier, and arrows mark intervals that extend below the plotted range. The number of catalogs in each tier is given below the panels.}
 \label{fig:tiers}
\end{figure*}
 
\begin{figure*}[t]
 \centering
 \includegraphics[width=\textwidth]{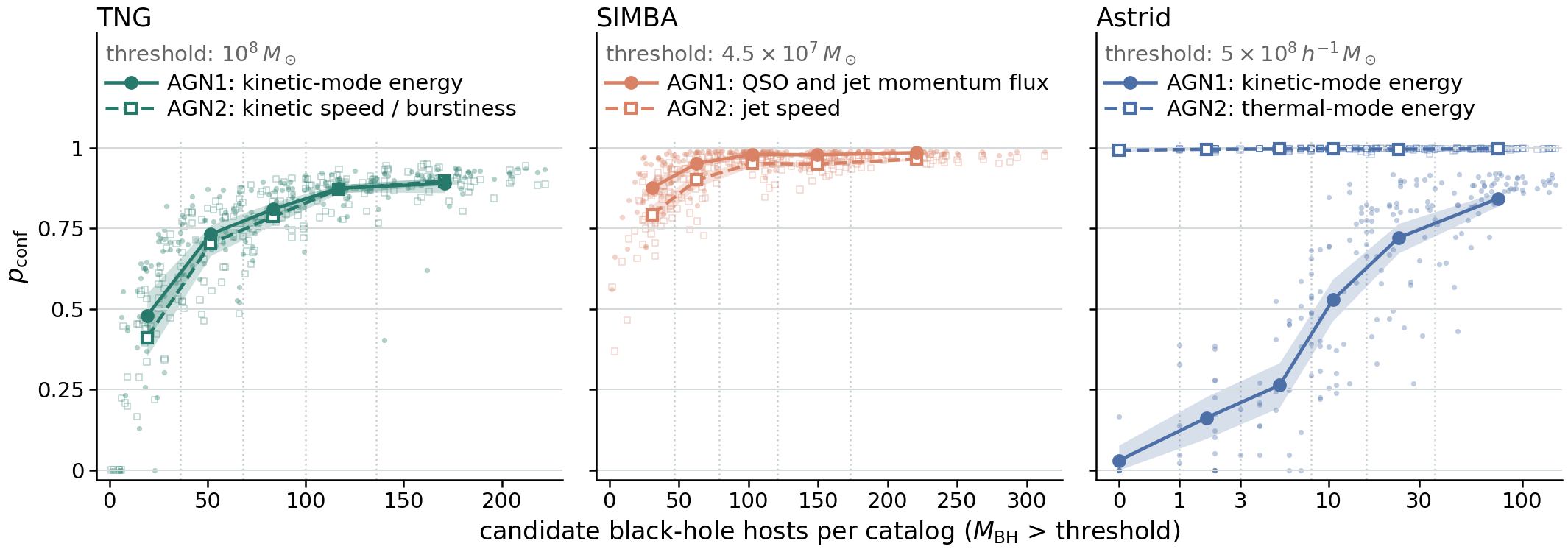}
 \caption{$p_{\mathrm{conf}}$ of the test catalogs for $A_{\mathrm{AGN1}}$ (circles, solid) and $A_{\mathrm{AGN2}}$ (squares, dashed) as a function of the number of subhalos in the catalog whose black-hole mass exceeds the threshold given in each panel. Centrals and satellites above the dark-matter threshold of Section~\ref{sec:data_catalogs} are counted. The Astrid axis is logarithmic in one plus this number. Small symbols show individual catalogs. Large symbols show the mean $p_{\mathrm{conf}}$ against the mean host count in each bin, with the bin edges marked by dotted lines, and shaded bands the 95\% bootstrap intervals of these means from 5000 resamples of the catalogs in each bin. The legend gives the physical process that each amplitude scales in each code, with QSO denoting the quasar mode.}
 \label{fig:pconf_bh}
\end{figure*}

The mean confidence over the test catalogs agrees with the accuracy of the readout. It agrees with the test $R^2$ to within 0.03 in 16 of the 18 combinations (Table~\ref{tab:readout}). The two exceptions are $A_{\mathrm{AGN2}}$ in IllustrisTNG and in SIMBA, where it is 0.73 and 0.91 against $R^2=0.63$ and 0.84. Because $p_{\mathrm{conf}}$ needs only the embedding, it anticipates how well each parameter is measured in a set of catalogs whose true parameters are unknown.
 
The score also ranks individual catalogs. We call a catalog high, middle or low confidence when $p_{\mathrm{conf}}\geq0.7$, $0.3\leq p_{\mathrm{conf}}<0.7$ or $p_{\mathrm{conf}}<0.3$. Figure~\ref{fig:tiers} compares the readout in the three tiers for the three parameters with the lowest overall $R^2$ ($A_{\mathrm{AGN1}}$ and $A_{\mathrm{AGN2}}$ in IllustrisTNG and $A_{\mathrm{AGN1}}$ in Astrid). For these parameters, 72.5\%, 69.5\% and 49\% of the test catalogs have high confidence. From the low to the high tier, $R^2$ rises and the RMSE falls in all three. The high-confidence catalogs are measured well, with $R^2=0.808$, 0.788 and 0.858 against 0.762, 0.627 and 0.615 for all test catalogs, and the RMSE drops from 0.176, 0.105 and 0.230 to 0.157, 0.081 and 0.145. In the low-confidence catalogs, the readout carries little information ($R^2=0.134$, $-0.627$ and $-0.090$).

For the well-measured AGN parameters, 97.5\% and 97\% of the SIMBA catalogs have high confidence for $A_{\mathrm{AGN1}}$ and $A_{\mathrm{AGN2}}$, and all Astrid catalogs have high confidence for $A_{\mathrm{AGN2}}$. The score thus picks out the catalogs whose AGN parameters are well measured.
 
The confidence follows the black-hole population of each catalog (Fig.~\ref{fig:pconf_bh}). Kinetic or jet feedback becomes important in black holes above $10^8\,\mathrm M_\odot$ in IllustrisTNG \citep{weinberger2017agn}, $4.5\times10^7\,\mathrm M_\odot$ in SIMBA \citep{ni2023camels} and $5\times10^8\,h^{-1}\,\mathrm M_\odot$ in Astrid \citep{ni2023camels}, and we count the subhalos whose black-hole mass exceeds these values. The median test catalog contains 103 such hosts in SIMBA, 82 in IllustrisTNG and 18 in Astrid, in the same order as the $R^2$ of $A_{\mathrm{AGN1}}$ (0.94, 0.76 and 0.62). In IllustrisTNG, the mean $p_{\mathrm{conf}}$ of $A_{\mathrm{AGN1}}$ and $A_{\mathrm{AGN2}}$ rises from 0.48 and 0.41 in catalogs with about 20 such hosts to 0.89 and 0.90 in catalogs with about 170. In SIMBA, catalogs with about 30 hosts already reach $p_{\mathrm{conf}}=0.88$ for $A_{\mathrm{AGN1}}$ and 0.79 for $A_{\mathrm{AGN2}}$, and above 100 hosts both exceed 0.9. In Astrid, $p_{\mathrm{conf}}$ for the kinetic-mode amplitude $A_{\mathrm{AGN1}}$ rises from 0.03 in catalogs without such black holes to 0.53 in catalogs with about 10 and 0.84 in catalogs with about 75. The thermal-mode amplitude $A_{\mathrm{AGN2}}$ of Astrid has $p_{\mathrm{conf}}>0.97$ in every catalog.
 
The information on kinetic and jet feedback in the embedding grows with the number of black holes massive enough to drive it. The error model is fitted to the readout residuals of the training catalogs and is never given the black-hole counts, and its confidence tracks this number from catalog to catalog.

\subsection{Statistical effect along the principal axes}
\label{sec:results_axes}
 
\begin{figure}[t]
 \centering
 \includegraphics[width=0.99\columnwidth]{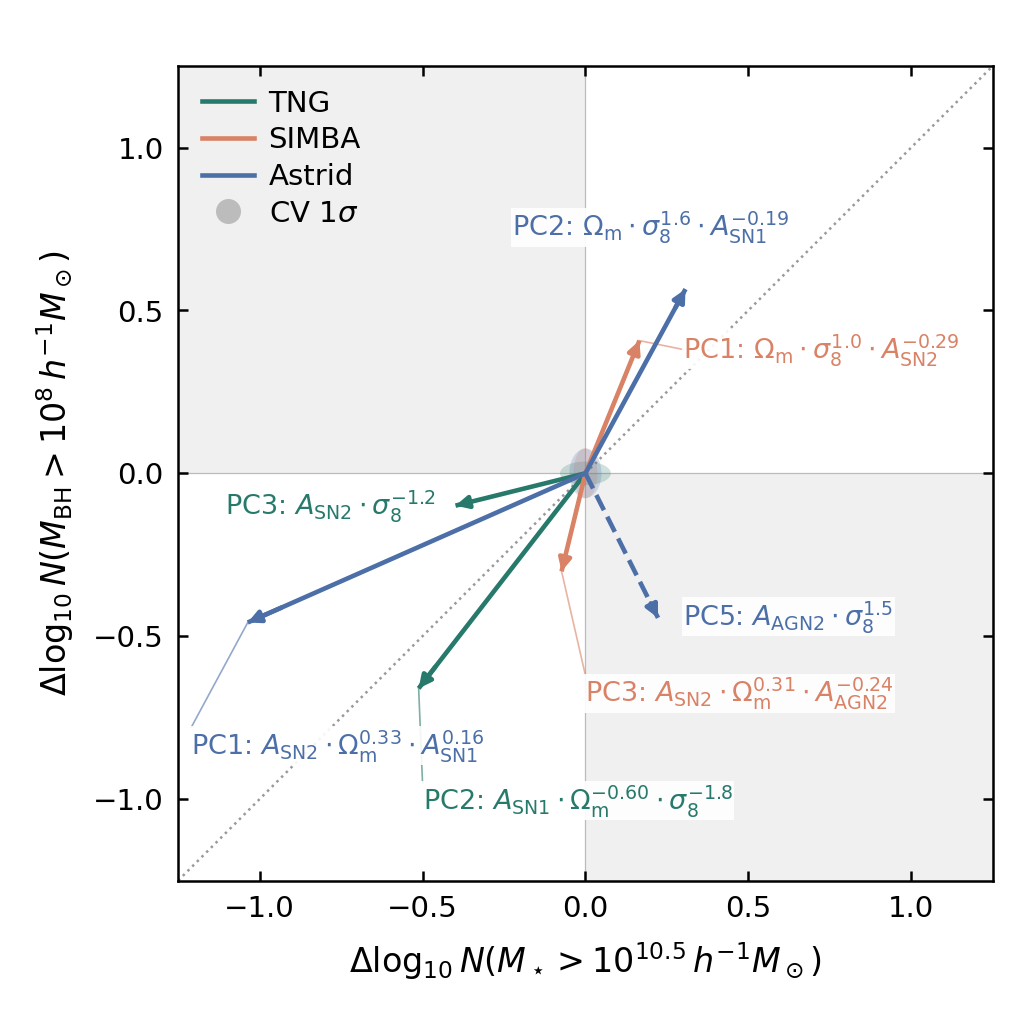}
 \caption{Changes in the numbers of massive galaxies and black holes along the principal axes of the embedding (Section~\ref{sec:results_axes}). Each arrow gives the change in $\log_{10}$ of the mean number of galaxies with $M_\star>10^{10.5}\,h^{-1}\,\mathrm M_\odot$ (horizontal) and of black holes with $M_{\mathrm{BH}}>10^{8}\,h^{-1}\,\mathrm M_\odot$ (vertical). Solid arrows show the two axes of each suite with the longest arrows, and the dashed arrow shows the axis of Astrid along which the two numbers change in opposite directions (shaded quadrants). Labels give the axis and the leading terms of its power law (equation~\ref{eq:axis_powerlaw}), and ellipses show the scatter between CV simulations.}
 \label{fig:axes}
\end{figure}

The directions studied so far are defined by the parameters. Here we use directions that the embedding defines without labels. For each suite, we compute $\boldsymbol z$ with $\mathcal T$ the 1,000 LH catalogs of the suite (equation~\ref{eq:catalog_pca}), so that its component $z_k$ is the coordinate of a catalog along the $k$-th principal axis of these catalogs. Along each axis, we take the points one standard deviation on either side of the mean and retrieve the five LH catalogs nearest to each point, as for Fig.~\ref{fig:interp}. The arrows in Fig.~\ref{fig:axes} give the change between the two sets of catalogs in $\log_{10}$ of the mean numbers of galaxies with $M_\star>10^{10.5}\,h^{-1}\,\mathrm M_\odot$ and of black holes with $M_{\mathrm{BH}}>10^{8}\,h^{-1}\,\mathrm M_\odot$. The figure shows the two axes of each suite with the longest arrows. The semi-axes of the ellipses are the standard deviations of the same changes between pairs of CV simulations of the suite, each represented by its five nearest LH catalogs, and measure the scatter from initial conditions.
 
The parameters enter only afterwards. Over the 1,000 LH catalogs, we fit the coordinate along each axis by least squares with a linear function of the logarithms of the six parameters $\theta_j$,
\begin{equation}
 \begin{split}
 z_k&\simeq c_k+\sum_{j=1}^{6}b_{kj}\log_{10}\theta_j\\
 &=c_k+b_{kd_k}\log_{10}\prod_{j=1}^{6}\theta_j^{\,b_{kj}/b_{kd_k}},
 \end{split}
 \label{eq:axis_powerlaw}
\end{equation}
which describes the axis by a power law of the parameters. Parameter $j$ accounts for the share $f_{kj}=b_{kj}^2\operatorname{Var}(\log_{10}\theta_j)/\operatorname{Var}(z_k)$ of the variance of $z_k$, and the leading parameter $\theta_{d_k}$ is the one with the largest share. Its exponent is one, and we orient each axis so that $b_{kd_k}>0$. The labels of Fig.~\ref{fig:axes} list the terms in decreasing order of $f_{kj}$ until they reach 95\% of $\sum_jf_{kj}$, with at most three terms, and each arrow points towards increasing power law. For the seven axes shown, the six-parameter fit accounts for at least 54\% of the variance of $z_k$.

Along most axes, massive galaxies and black holes change in the same direction. In IllustrisTNG and Astrid, the axes that follow the supernova parameters lower both. The leading axis of Astrid follows mainly the wind speed $A_{\mathrm{SN2}}$ and reduces the number of massive galaxies by a factor of 11 and that of massive black holes by a factor of 2.9. In IllustrisTNG, the second axis follows $A_{\mathrm{SN1}}\Omega_{\mathrm m}^{-0.60}\sigma_8^{-1.8}$ and reduces them by factors of 3.2 and 4.6, and the third axis follows $A_{\mathrm{SN2}}\sigma_8^{-1.2}$ and reduces them by factors of 2.5 and 1.3.
 
In SIMBA, the third axis follows mainly $A_{\mathrm{SN2}}$ and reduces the number of massive black holes by a factor of 2.0 and that of massive galaxies by 1.2, in line with the suppression of black-hole growth by faster winds (Section~\ref{sec:results_response}). The leading axis of SIMBA and the second axis of Astrid follow $\Omega_{\mathrm m}$ and $\sigma_8$. They raise the numbers of massive black holes by factors of 2.6 and 3.7 and those of massive galaxies by 1.5 and 2.0.
 
One axis moves the two populations apart. Among the axes of all three suites whose six-parameter fit accounts for at least half of the variance of $z_k$, only the fifth axis of Astrid changes the two numbers in opposite directions, each by more than three times the corresponding semi-axis of the ellipse. Figure~\ref{fig:axes} shows it as a dashed arrow. This axis follows $A_{\mathrm{AGN2}}\sigma_8^{1.5}$, with the thermal-mode amplitude as its leading term. Along it, massive galaxies become 1.7 times more numerous and massive black holes 2.8 times rarer. The same effect appears as the anti-alignment of the thermal and kinetic AGN directions (Section~\ref{sec:results_response}) and as the lower black-hole masses along $A_{\mathrm{AGN2}}$ in Fig.~\ref{fig:interp}. In the picture of self-regulated growth of Section~\ref{sec:results_response}, stronger thermal feedback leaves fewer black holes massive enough to enter the kinetic mode, and fewer massive galaxies are then quenched by it.
 
The principal axes of the embedding follow combinations of parameters that change the numbers of massive galaxies and black holes. Among the axes shown, the number of massive galaxies changes most along the supernova axes in IllustrisTNG and Astrid and along the cosmological axis in SIMBA.

%% file: 05-Discussion.tex
\section{Discussion}
\label{sec:discussion}
This section discusses what the embedding shares across galaxy-formation models (Section~\ref{sec:disc_shared}), where its information on the parameters comes from (Section~\ref{sec:disc_objective}), how it can be used (Section~\ref{sec:disc_uses}) and how it can be extended to observations and to broader simulation suites (Section~\ref{sec:disc_limits}).

\subsection{A shared representation across galaxy-formation models}
\label{sec:disc_shared}

The embedding keeps the catalogs of the three codes in separate regions (Fig.~\ref{fig:catalog_pca}), and a change of cosmology moves them in partly aligned directions (Fig.~\ref{fig:similarity}). One representation holds both the differences between the codes and a shared response to cosmology, and this shared response carries over to Swift-EAGLE, which the encoder never saw (Section~\ref{sec:results_swift}). Supervised networks infer $\Omega_{\mathrm m}$ robustly across codes from the positions and velocities of galaxies or halos \citep{desanti2023robust,shao2023halos}, and internal properties can tie them more closely to the code they were trained on \citep{villaescusa2022onegalaxy,echeverri2023onegalaxy,shao2023halos}. The zero-shot readout of Celia draws on the internal properties as well and still follows $\Omega_{\mathrm m}$ in the new code. The relations that the self-supervised objective learns between the dark-matter and baryonic properties of subhalos appear to be partly shared among galaxy-formation models.

Each feedback amplitude rescales a process of one galaxy-formation model relative to the fiducial value of that model, so the same name can stand for different physics in different codes (Table~\ref{tab:data_suites}). Readouts of these amplitudes use labels from the new code, and a small labeled set suffices (Fig.~\ref{fig:label_efficiency}). The encoder itself stays unchanged, so adding a code requires a small labeled set and no retraining.

Parameters that scale similar processes in different codes also move the catalogs in similar directions (Section~\ref{sec:results_response}), and most of these alignments match the responses of galaxies, black holes and gas in the 1P sets \citep{delgado2023clustering,ni2023camels,gebhardt2024spread}. They are candidates for feedback coordinates that are defined across codes. An encoder with one part shared by the codes and another specific to each code would make such coordinates explicit and could carry more of the readout calibration across codes. \citet{mercader2026disentangling} learn such a split from overlapping observations of the same galaxies, separating the physical content of the galaxies from the effects of each survey. In simulations, the code-specific part would describe the galaxy-formation model of each code and would itself be a subject of study. Simulations that share initial conditions and cosmology and differ only in their galaxy-formation code would provide the training pairs, and the 1P sets of IllustrisTNG, SIMBA and Astrid already share their initial random seed \citep{villaescusa2021camels,ni2023camels}. Once calibrated against quantities that all codes predict, such as the baryon fractions of halos or the abundance of massive black holes, the shared feedback directions would become targets for inference across codes.

\subsection{Sources of parameter information}
\label{sec:disc_objective}

Masked reconstruction and the prediction of the teacher's object representations both ask the encoder to infer the hidden properties of a subhalo from its visible properties and from the objects around it (Section~\ref{sec:objectives}). Relations such as those between the dark-matter and stellar masses of a subhalo and between a satellite and its host change with cosmology and with the galaxy-formation physics of the simulation. The visible objects of a view carry the imprint of these parameters, so the objective rewards an encoder that uses this imprint to predict the hidden properties. The covariance of the object representations collects these relations over a whole catalog and retains the parameters, which the encoder never received (Section~\ref{sec:results_readout}).

Much of the information on $\Omega_{\mathrm m}$ comes from the internal properties of the subhalos (Section~\ref{sec:results_restricted}). The maximum circular velocity traces the dark matter of a galaxy, and the objective asks the encoder to predict the dark-matter properties of a subhalo from its baryonic ones and the reverse. CAMELS fixes $\Omega_{\mathrm b}$, so $\Omega_{\mathrm m}$ sets the dark-matter content of galaxies relative to their baryons \citep{villaescusa2022onegalaxy}, and the baryon fractions of halos carry $\Omega_{\mathrm m}$ directly. Part of the precision on $\Omega_{\mathrm m}$ therefore reflects the design of the simulations.

With positions alone, the readout stays below the GNNs of \citet{villanueva2022cosmicgraphs}. This input lies outside the training conditions of the encoder. Its training views hide random groups of properties and random objects, and most objects keep some of their properties, so the encoder rarely sees a catalog in which every object is reduced to its position. The GNNs are built on graphs of paired relations and learn the correlations between positions directly. \citet{baldwin2026inductivebiases} find that GNNs encoding galaxy--galaxy relations make the best use of galaxy positions, and that Deep Sets draw the information on $\Omega_{\mathrm m}$ mainly from the line-of-sight velocities. Training views in which every object keeps only its position, or a reconstruction target for the spatial distribution of the objects, would ask the encoder for this clustering information as well.

The kinetic-mode amplitudes of IllustrisTNG and Astrid are the hardest parameters to measure. They act directly on the few subhalos whose black holes are massive enough to drive kinetic feedback, and the confidence of their readouts grows with the number of such hosts (Section~\ref{sec:results_pconf}). The covariance of equation~(\ref{eq:catalog_covariance}) weights every subhalo equally, and mass-weighted pooling or attention pooling over the most massive halos would give these hosts more weight. At fixed parameters, the number of hosts grows in proportion to the volume, and a box with four times the CAMELS volume would contain about as many hosts as the Astrid catalogs with the highest confidence. Simulations of this size would measure how the information on kinetic feedback grows with the number of hosts. The label-efficiency curves of these parameters still rise at the largest training set (Fig.~\ref{fig:label_efficiency}), so more labeled simulations would raise their precision as well.

\subsection{Uses of the embedding}
\label{sec:disc_uses}

In astronomy, foundation models such as AstroCLIP \citep{parker2024astroclip} and AION-1 \citep{parker2025aion} learn embeddings of galaxy images and spectra from which simple heads recover physical properties, redshifts and classifications. Celia plays the same role for galaxy catalogs. It is trained once without labels, and each downstream task fits only a readout on the frozen embedding of the whole catalog.

Before any readout is fitted, the geometry of the embedding shows which parameters move the catalogs of a suite beyond the scatter from initial conditions, and hence which parameters the suite can constrain at its volume. In CAMELS, the kinetic-mode amplitudes of IllustrisTNG and Astrid move a catalog least, and the direction of $A_{\mathrm{AGN1}}$ in IllustrisTNG overlaps with those of the cosmological parameters and of the initial conditions (Section~\ref{sec:results_response}). Once a readout and its error model are fitted, the confidence score extends this diagnosis to single catalogs. Its mean follows the readout $R^2$ because the error model is calibrated on the training catalogs, and its variation from catalog to catalog picks out the catalogs whose AGN parameters are well measured (Section~\ref{sec:results_pconf}).

The principal axes of the embedding and the retrieval of the nearest LH catalogs let one explore a simulation suite without choosing a summary statistic in advance. The axes are found from the catalogs alone, and the statistics and the parameters enter only afterwards, so the same axes can be read with any quantity that the catalogs provide. In Astrid, this exploration finds an axis led by the thermal-mode amplitude along which massive black holes and massive galaxies change in opposite directions (Section~\ref{sec:results_axes}). This matches the response of the Astrid 1P set \citep{ni2023camels} and the self-regulated growth of black holes \citep{booth2009agn}. Axes that follow the cosmological parameters recur in SIMBA and Astrid, and such recurring axes could guide the choice of parameters for new simulation suites. Retrieval also turns any direction in the embedding into a sequence of simulated catalogs (Fig.~\ref{fig:interp}).

The embedding can also enter inference and emulation directly. Its readout predictions are compressed summaries of the kind used to make likelihood-free inference tractable \citep{alsing2018compression}, and a neural posterior estimator trained on the embedding would give the joint posterior of several parameters \citep{cranmer2020sbi} and may recover more of the AGN information than a linear map. The 1P sets trace the response of the embedding to each parameter with the random seed of the initial conditions held fixed, and the CV sets measure the scatter that the initial conditions add (Section~\ref{sec:results_response}). A conditional density model of $\boldsymbol z$ given the parameters can be trained to reproduce both. Such a model would emulate the embeddings of new simulations without running them, and the linear readouts would apply to its samples unchanged.

\subsection{Observations and broader simulation suites}
\label{sec:disc_limits}

The catalogs used here are simulated at $z=0$, with real-space positions, full three-dimensional velocities and properties that are not directly observable, such as dark-matter masses and $V_{\max}$. Applying the encoder to galaxy surveys requires inputs restricted to observables in redshift space, with survey selection and measurement noise. The frozen encoder already embeds restricted inputs without retraining, and with line-of-sight velocities or stellar properties its readouts reach a precision comparable to that of supervised networks (Section~\ref{sec:results_restricted}). Training views with survey selection and redshift-space distortions would extend the encoder to such data, and catalogs at higher redshift can enter the training as additional views.

The encoder was trained on the six-parameter Latin hypercubes of three codes. The SB28 set of IllustrisTNG varies 28 parameters at once, and the 1P-28 sets of IllustrisTNG and SIMBA vary the additional parameters one at a time \citep{ni2023camels}. These sets would test whether the shared directions persist as more physical processes vary. Additional codes in the pretraining would show whether the zero-shot transfer improves with the number of codes. A generative decoder from the embedding to catalogs or their statistics would extend the retrieval of Sections~\ref{sec:results_response} and~\ref{sec:results_axes} beyond the simulated catalogs.

%% file: 06-Conclusion.tex
\section{Conclusions}
\label{sec:conclusions}

We have presented Celia, a self-supervised encoder trained without labels on the subhalo catalogs of the IllustrisTNG, SIMBA and Astrid suites of CAMELS and their N-body counterparts. During training, the encoder fills in subhalo properties hidden from a masked view and predicts the representations that an EMA teacher assigns to the same subhalos in the complete view. The frozen encoder then turns each catalog into one vector. Linear readouts of this vector estimate the simulation parameters with error bars. The same vectors define a response direction for each parameter, and the nearest simulated catalogs translate directions in the embedding into catalog statistics. Our main findings are as follows.

\begin{enumerate}
    \item The embedding retains the cosmological and supernova parameters almost completely. In all three suites, linear readouts reach $R^2=0.999$ for $\Omega_{\mathrm m}$, 0.986 to 0.991 for $\sigma_8$ and at least 0.97 for both supernova amplitudes. The precision for the AGN amplitudes depends on how many objects they act on. In SIMBA they reach $R^2=0.94$ and 0.84, the thermal mode of Astrid reaches 0.996, and the kinetic-mode amplitudes of IllustrisTNG and Astrid reach 0.62 to 0.76. With the restricted inputs of earlier supervised studies, the readouts are comparable to their graph neural networks once the positions are complemented by line-of-sight velocities or stellar properties.
    \item Celia transfers to Swift-EAGLE without retraining. The readout fitted on the three training suites follows the Swift-EAGLE values of $\Omega_{\mathrm m}$ with $R^2=0.825$ and $r=0.985$, and those of $\sigma_8$ with $r=0.79$. The feedback amplitudes refer to the model of each code and need labels from the new code. Readouts fitted on 20 labeled Swift-EAGLE catalogs give $r\geq0.94$ for $\Omega_{\mathrm m}$, $\sigma_8$ and $A_{\mathrm{SN1}}$, and readouts fitted on 100 give $r\geq0.96$ for every parameter except $A_{\mathrm{AGN2}}$ ($r=0.85$), similar to the number of labels that the training codes need.
    \item The geometry of the embedding separates the effect of each parameter from the scatter of the initial conditions. The cosmological and supernova parameters move a catalog by 4 to 59 times this scatter, and the kinetic-mode amplitudes move it least. Feedback parameters that scale similar processes in different codes point in similar directions, and most of these alignments match the responses of galaxies, black holes and gas in the 1P simulations \citep{delgado2023clustering,ni2023camels,gebhardt2024spread}. Simulations retrieved along straight lines in the embedding show the changes in their statistics that each parameter is expected to cause.
    \item The confidence score of each prediction comes from the embedding alone, and for the AGN parameters it closely follows the massive black-hole content of the catalog. For the kinetic- and jet-mode amplitudes, it rises with the number of black holes massive enough to drive that feedback. For the kinetic mode of Astrid, its mean grows from 0.03 in catalogs without such black holes to 0.84 in catalogs with about 75.
    \item The principal axes of the embedding need no labels. Each follows a combination of parameters, and moving along it changes the numbers of massive galaxies and black holes. In Astrid, the axis led by the thermal-mode amplitude raises the number of massive galaxies by a factor of 1.7 and lowers that of massive black holes by a factor of 2.8, matching the response of the Astrid 1P set to stronger thermal feedback \citep{ni2023camels}. In SIMBA and Astrid, the axes that follow $\Omega_{\mathrm m}$ and $\sigma_8$ raise the numbers of both populations.
\end{enumerate}

Celia provides a shared coordinate system for simulated universes. Catalogs from different galaxy-formation models can be compared in it, and its frozen encoder places the catalogs of a new code there without retraining. Inference, diagnostics and the exploration of a simulation suite all operate on one vector per catalog. Larger simulation volumes, survey-like training views and more galaxy-formation models would extend this coordinate system toward observed galaxy catalogs.

%% file: 07-Appendix.tex
\section{Input fields and training settings}
\label{app:details}

\paragraph{Input fields.} Tables~\ref{tab:catalog_features} and~\ref{tab:catalog_features_host} list the 53 input features of a subhalo and their transformations before standardization. Masses, radii, rates and characteristic velocities such as $V_{\max}$ enter as base-10 logarithms in catalog units. These units are $10^{10}\,h^{-1}\,\mathrm M_\odot$ for masses, $h^{-1}\,\mathrm{ckpc}$ for lengths, $\mathrm{km\,s^{-1}}$ for velocities and $\mathrm M_\odot\,\mathrm{yr}^{-1}$ for star-formation rates. Positions are fractional box coordinates $\boldsymbol x=(\boldsymbol r\bmod L)/L$, and all separations use the minimum image in the periodic box. Positions and velocities relative to the host are normalized by the $R_{200\mathrm c}$ and $V_{200\mathrm c}$ of the host, and the center offsets of the subhalo and of its host by the total half-mass radius and by $R_{200\mathrm c}$, respectively. The host gas, stellar and black-hole masses enter as fractions of the host FoF mass, and the host dark-matter fraction as one minus the summed mass fractions of all other particle types.

All features other than the positions are standardized with means and standard deviations computed on the training catalogs. The three components of the velocity and of the host-relative position and velocity share one scale and have zero mean. The Swift-EAGLE catalogs are standardized with the same statistics. The positivity indicators enter the encoder standardized in the same way and serve as reconstruction targets with their original values 0 and 1.

The inputs exclude the particle counts of subhalos and groups, because together with the masses they give the particle mass and hence $\Omega_{\mathrm m}$. They also exclude the number of subhalos in a group and the catalog radii $R_{200\mathrm c}$ and $R_{500\mathrm c}$, which follow from the unperturbed masses. The subhalo total mass is not an input either (Section~\ref{sec:training_masks}). The spin enters only through its magnitude, since its direction is statistically isotropic.

\paragraph{Mass perturbation.} The perturbation of equation~(\ref{eq:data_mass_perturbation}) shifts the dark-matter mass of each subhalo by an offset $\delta_M$ and the three host masses $M_{\mathrm{FoF}}$, $M_{200\mathrm c}$ and $M_{500\mathrm c}$ by a common offset $\delta_{M,\mathrm h}$. The host scales
\begin{equation}
 R'_{200\mathrm c}=\left(\frac{3M'_{200\mathrm c}}{800\pi\rho_{\mathrm{crit},0}}\right)^{1/3},\qquad
 V'_{200\mathrm c}=\left(\frac{GM'_{200\mathrm c}}{R'_{200\mathrm c}}\right)^{1/2}
 \label{eq:host_scales}
\end{equation}
are computed from the perturbed masses, and so are the host fractions and the quantities normalized by these scales. No combination of inputs therefore returns an unperturbed mass.

\paragraph{Training settings.} We optimize the student with AdamW with $\beta_1=0.9$, $\beta_2=0.95$ and weight decay 0.05 and clip the gradient norm at 10. The weight decay does not act on one-dimensional parameters such as biases and normalization weights. The forward pass uses bfloat16 mixed precision, and the reconstruction loss is computed in single precision. An epoch has 63 batches. The learning rate rises linearly to $2\times10^{-4}$ over the first 200 steps, stays constant until step 10,000 and then decays linearly to zero at step 18,000. The EMA momentum is $m=0.996$ during the warmup and the stable phase and rises to 1 during the decay,
\begin{equation}
 m(t)=1-\frac{1-0.996}{2}\left[1+\cos\!\left(\pi\,\frac{t-10^4}{8\times10^3}\right)\right],
 \label{eq:ema_schedule}
\end{equation}
for $10^4\leq t\leq1.8\times10^4$.

\begin{table*}
 \centering
 \caption{Subhalo input fields and their transformations before standardization. Dim.\ gives the number of input features of each field, including its positivity indicator where present, and the last column gives the input group of each field (Appendix~\ref{app:groups}). Together with Table~\ref{tab:catalog_features_host}, a subhalo has 53 input features. The positions enter only the spatial modules. Here $q$ is the field value in catalog units, $\boldsymbol r$ a position, $\boldsymbol v$ a velocity and $\delta_M$ the mass offset of equation~(\ref{eq:data_mass_perturbation}).}
 \label{tab:catalog_features}
 \footnotesize
 \setlength{\tabcolsep}{3pt}
 \renewcommand{\arraystretch}{1.10}
  \begin{tabular}{@{}lllll@{}}
  \toprule
  Catalog field & Dim. & Quantity & Transformation & Input group\\
  \midrule
  \multicolumn{5}{@{}l}{\emph{Geometry and kinematics}}\\
  \texttt{SubhaloPos} & 3 & Position & $(\boldsymbol r\bmod L)/L$ & ---\\
  \texttt{SubhaloVel} & 3 & Velocity & $\boldsymbol v/10^3$ & velocity\\
  \midrule
  \multicolumn{5}{@{}l}{\emph{Subhalo dark matter and structure}}\\
  \texttt{SubhaloMassType[1]} & 1 & Dark-matter mass & $\log_{10}q+\delta_M$ & DM mass\\
  \texttt{SubhaloHalfmassRadType[1]} & 1 & DM half-mass radius & $\log_{10}q$ & half-mass radii\\
  \texttt{SubhaloHalfmassRad} & 1 & Total half-mass radius $R_{1/2}$ & $\log_{10}q$ & half-mass radii\\
  \texttt{SubhaloVmax} & 1 & Maximum circular velocity $V_{\max}$ & $\log_{10}q$ & dynamics\\
  \texttt{SubhaloVmaxRad} & 1 & Radius of $V_{\max}$ & $\log_{10}q$ & dynamics\\
  \texttt{SubhaloVelDisp} & 1 & Velocity dispersion & $\log_{10}q$ & dynamics\\
  $\lVert\texttt{SubhaloSpin}\rVert$ & 1 & Specific angular momentum & $\log_{10}q$ & dynamics\\
  $\lVert\texttt{SubhaloPos}-\texttt{SubhaloCM}\rVert$ & 1 & Center offset $d$ & $\ln(1+d/R_{1/2})$ & center offset\\
  \midrule
  \multicolumn{5}{@{}l}{\emph{Subhalo baryons}}\\
  \texttt{SubhaloMassType[4]} & 2 & Stellar mass & $\log_{10}q$, $\mathbf 1_{q>0}$ & stellar\\
  \texttt{SubhaloHalfmassRadType[4]} & 1 & Stellar half-mass radius $R_{1/2,\star}$ & $\log_{10}q$ & stellar radius\\
  \texttt{SubhaloStarMetallicity} & 2 & Stellar metallicity within $2R_{1/2,\star}$ & $\log_{10}q$, $\mathbf 1_{q>0}$ & stellar metallicity\\
  \texttt{SubhaloMassType[0]} & 2 & Gas mass & $\log_{10}q$, $\mathbf 1_{q>0}$ & gas\\
  \texttt{SubhaloGasMetallicity} & 2 & Gas metallicity within $2R_{1/2,\star}$ & $\log_{10}q$, $\mathbf 1_{q>0}$ & gas metallicity\\
  \texttt{SubhaloGasMetallicitySfr} & 2 & Metallicity of star-forming gas & $\log_{10}q$, $\mathbf 1_{q>0}$ & SF-gas metallicity\\
  \texttt{SubhaloSFR} & 2 & Star-formation rate & $\log_{10}q$, $\mathbf 1_{q>0}$ & star formation\\
  \texttt{SubhaloSFRinRad} & 2 & SFR within $2R_{1/2,\star}$ & $\log_{10}q$, $\mathbf 1_{q>0}$ & star formation\\
  \midrule
  \multicolumn{5}{@{}l}{\emph{Subhalo black holes}}\\
  \texttt{SubhaloBHMass} & 2 & Black-hole mass & $\log_{10}q$, $\mathbf 1_{q>0}$ & black hole\\
  \texttt{SubhaloBHMdot} & 2 & Black-hole accretion rate & $\log_{10}q$, $\mathbf 1_{q>0}$ & BH accretion\\
  \bottomrule
 \end{tabular}
\end{table*}

\begin{table*}
 \centering
 \caption{Input fields of the host group and of the subhalo relative to its host, continuing Table~\ref{tab:catalog_features}. Here $q$ is the field value in catalog units, $M_{\mathrm h,k}$ denotes \texttt{GroupMassType[}$k$\texttt{]}, $\Delta\boldsymbol r$ and $\Delta\boldsymbol v$ are the position and velocity relative to the host, $\delta_{M,\mathrm h}$ is the host-mass offset, $f'_{\mathrm{DM}}=1-\sum_{k\neq1}M_{\mathrm h,k}/M'_{\mathrm{FoF}}$, and primes mark quantities computed from perturbed masses (equation~\ref{eq:host_scales}).}
 \label{tab:catalog_features_host}
 \footnotesize
 \setlength{\tabcolsep}{3pt}
 \renewcommand{\arraystretch}{1.10}
  \begin{tabular}{@{}lllll@{}}
  \toprule
  Catalog field & Dim. & Quantity & Transformation & Input group\\
  \midrule
  $\sum_k\texttt{GroupMassType[}k\texttt{]}$ & 1 & FoF mass $M_{\mathrm{FoF}}$ & $\log_{10}q+\delta_{M,\mathrm h}$ & FoF mass\\
  \texttt{Group\_M\_Crit200} & 1 & Halo mass $M_{200\mathrm c}$ & $\log_{10}q+\delta_{M,\mathrm h}$ & $M_{200\mathrm c}$\\
  \texttt{Group\_M\_Crit500} & 1 & Halo mass $M_{500\mathrm c}$ & $\log_{10}q+\delta_{M,\mathrm h}$ & $M_{500\mathrm c}$\\
  $\lVert\texttt{GroupPos}-\texttt{GroupCM}\rVert$ & 1 & Center offset $d_{\mathrm h}$ & $\ln(1+d_{\mathrm h}/R'_{200\mathrm c})$ & host center offset\\
  \texttt{GroupMassType[}$k\neq1$\texttt{]} & 1 & Dark-matter fraction $f'_{\mathrm{DM}}$ & $\log_{10}f'_{\mathrm{DM}}$ & host DM fraction\\
  \texttt{GroupMassType[0]} & 2 & Gas fraction & $\log_{10}(q/M'_{\mathrm{FoF}})$, $\mathbf 1_{q>0}$ & gas\\
  \texttt{GroupMassType[4]} & 2 & Stellar fraction & $\log_{10}(q/M'_{\mathrm{FoF}})$, $\mathbf 1_{q>0}$ & stellar\\
  \texttt{GroupBHMass} & 2 & Black-hole mass ratio & $\log_{10}(q/M'_{\mathrm{FoF}})$, $\mathbf 1_{q>0}$ & black hole\\
  \texttt{GroupSFR} & 2 & Star-formation rate & $\log_{10}q$, $\mathbf 1_{q>0}$ & star formation\\
  \texttt{SubhaloPos} $-$ \texttt{GroupPos} & 3 & Position relative to the host & $\Delta\boldsymbol r/R'_{200\mathrm c}$ & host-relative\\
  \texttt{SubhaloVel} $-$ \texttt{GroupVel} & 3 & Velocity relative to the host & $\Delta\boldsymbol v/V'_{200\mathrm c}$ & host-rel.\ velocity\\
  \texttt{GroupFirstSub} & 1 & Central flag & $\mathbf 1_{q=i}$, subhalo index $i$ & host-relative\\
  \bottomrule
 \end{tabular}
\end{table*}

\section{Encoder and masking details}
\label{app:encoder_training}

This appendix specifies the input groups and masking families (Appendix~\ref{app:groups}), the spatial modules (Appendix~\ref{app:spatial}), the propagation of masks between related fields (Appendix~\ref{app:masking}) and the reconstruction targets (Appendix~\ref{app:reconstruction}).

\subsection{Input groups and masking families}
\label{app:groups}

The last columns of Tables~\ref{tab:catalog_features} and~\ref{tab:catalog_features_host} assign every field other than the positions to one of the 21 input groups of equation~(\ref{eq:tokenization}). Several groups join a subhalo field with the corresponding host field. The stellar, gas, star-formation and black-hole groups also contain the host stellar, gas and black-hole mass fractions and the host star-formation rate, so step~1 of the masking hides these host fields as well.

In terms of these groups, the four baryonic families are \{stellar, stellar radius, stellar metallicity\}, \{gas, gas metallicity\}, \{star formation, SF-gas metallicity\} and \{black hole, BH accretion\}. The seven non-baryonic families are \{velocity, host-rel.\ velocity\}, \{FoF mass, $M_{200\mathrm c}$, $M_{500\mathrm c}$\}, \{DM mass\}, \{half-mass radii\}, \{dynamics\}, \{center offset\} and \{host center offset\}. The host-relative group holds the host-relative positions and the central flag. This group and the host DM fraction belong to no family.

\subsection{Spatial modules}
\label{app:spatial}

The pair-attention layer of equation~(\ref{eq:spatial_attention}) has eight heads of dimension 32. One MLP takes the three separation measures as input and returns the eight biases $B_h$. Its output layer is initialized to zero, so that the layer starts as unbiased attention. Each spatial module adds its output to the tokens and is followed by a residual feed-forward block with pre-layer normalization, a hidden width of 1,024 and GELU activation.

\subsection{Mask propagation}
\label{app:masking}

Choosing whole hosts in step~3 hides every subhalo with probability 0.2, whatever the size of its host. Step~4 hides fields that could be computed from visible ones. The host dark-matter fraction equals one minus the summed fractions of the other particle types (Table~\ref{tab:catalog_features_host}), so it is hidden whenever the host gas, stellar or black-hole fraction is hidden.

The host-relative positions $\Delta\tilde{\boldsymbol r}=\Delta\boldsymbol r/R'_{200\mathrm c}$ and velocities $\Delta\tilde{\boldsymbol v}=\Delta\boldsymbol v/V'_{200\mathrm c}$ carry the scales of the host. The positions are always visible, and for two members $a$ and $b$ of the same host they give
\begin{equation}
 R'_{200\mathrm c}=\frac{\lVert\boldsymbol r_a-\boldsymbol r_b\rVert}{\lVert\Delta\tilde{\boldsymbol r}_a-\Delta\tilde{\boldsymbol r}_b\rVert},
 \label{eq:r200_recovery}
\end{equation}
and the velocities give $V'_{200\mathrm c}$ in the same way whenever the host-relative velocities are visible. Either scale returns $M'_{200\mathrm c}$ through equation~(\ref{eq:host_scales}). Whenever the host masses are hidden, the host-relative positions and velocities, the host center offset and the central flag are therefore hidden for all members of the host. This rule acts on single groups, so it can hide the host-relative velocities and leave the velocities of the same family visible.

\subsection{Reconstruction targets}
\label{app:reconstruction}

Binary targets that a visible indicator already determines are not scored. These are a positive gas mass when the star-formation rate is visible and positive, a positive stellar mass or host stellar fraction when the gas metallicity or the star-formation rate within $2R_{1/2,\star}$ is visible and positive, and a positive host gas fraction when the star-formation rate of the subhalo or of its host is visible and positive.

\section{Readout details}
\label{app:readout_details}
 
With $d_{\mathcal T}$ retained principal components of variances $\lambda_1,\dots,\lambda_{d_{\mathcal T}}$ over $\mathcal T$, the scalar of equation~(\ref{eq:catalog_pca}) is
\begin{equation}
 s_{\mathcal T}=\Bigl(\frac{1}{d_{\mathcal T}}\sum_{j=1}^{d_{\mathcal T}}\lambda_j\Bigr)^{-1/2}.
 \label{eq:pca_scale}
\end{equation}
It keeps the relative variances of the components. The projection and the scalar are fitted on $\mathcal T$ and applied unchanged to the test catalogs.
 
The regularization strength $\alpha$ is chosen separately for each target from 33 logarithmically spaced values between $10^{-3}$ and $10^5$. The training catalogs are split by their LH index into three folds. Every suite uses the same assignment of LH indices to folds, so catalogs with the same LH index fall into the same fold of a pooled readout. Test catalogs are predicted by the readout fitted on all training catalogs, with $\alpha$ minimizing the cross-validated squared error over the three folds. Each training catalog also receives an out-of-fold prediction $\widehat y^{\,\mathrm{oof}}_n$ from a readout fitted on the other two folds, with $\alpha$ chosen by a three-fold cross-validation within these catalogs.
 
The error model is a ridge regression with an unpenalized intercept, and its regularization strength is chosen on the same grid and folds as for the readout. For the multipliers, the error model of each fold computes its floor $\eta$ and the variance $V_0$ on the other two folds.

\section{Calibration of the error bars}
\label{app:pit}
 
\begin{figure*}[t]
 \centering
 \includegraphics[width=\textwidth]{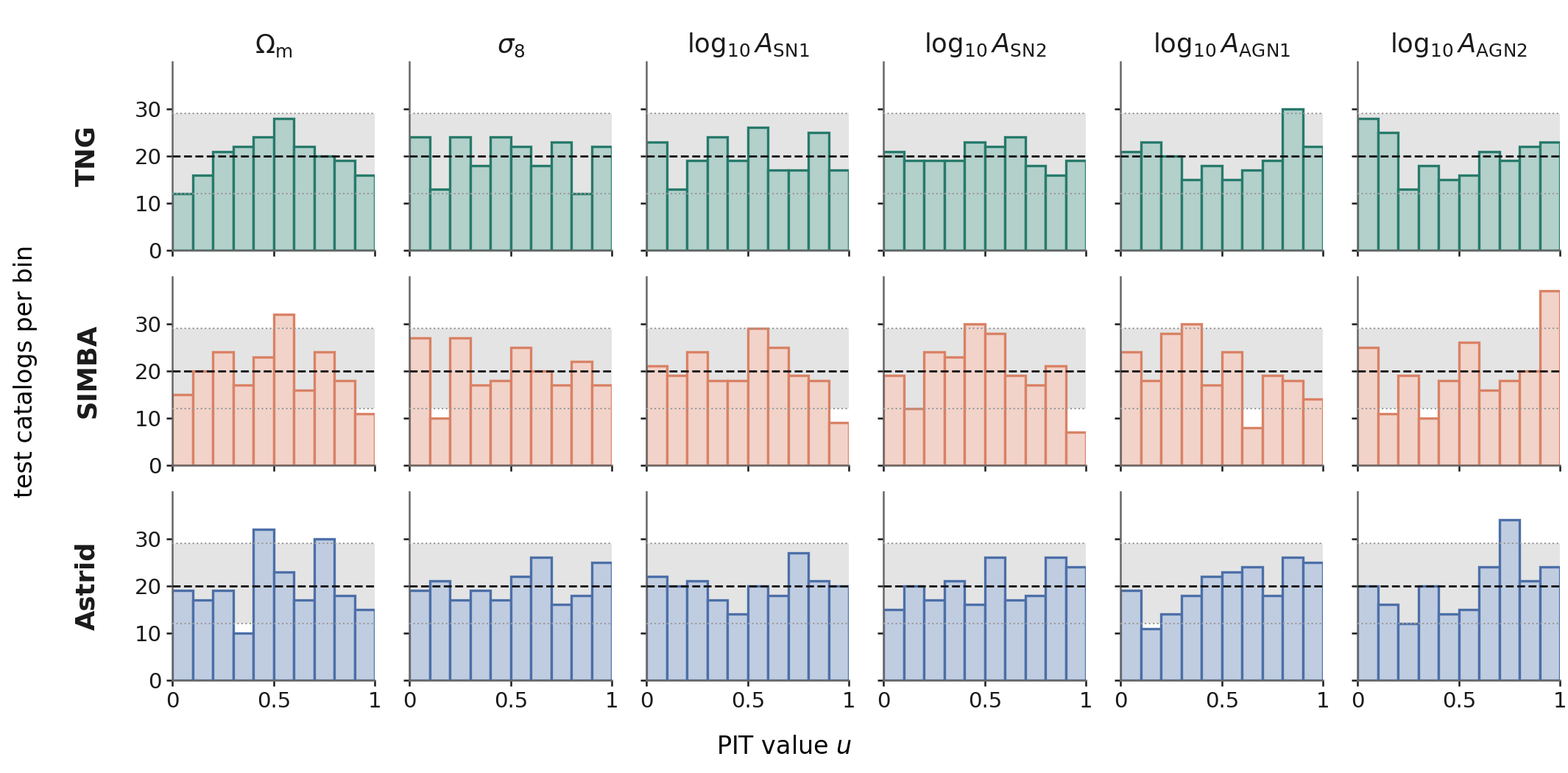}
 \caption{Calibration of the error bars on the 200 test catalogs of each suite (rows). Each histogram shows the probability integral transform $u$ of the true values under the predictive distribution of the readout, in ten bins. The dashed line marks the expected 20 catalogs per bin, and the gray band the central 95\% range of the counts for uniform $u$.}
 \label{fig:pit}
\end{figure*}
 
For each test catalog and parameter, the probability integral transform (PIT) $u$ is the fraction of training catalogs whose signed normalized out-of-fold residual $r_n/\widehat\sigma^{\,\mathrm{oof}}_n$ lies below $(y-\widehat y)/\widehat\sigma(\boldsymbol z)$ (Section~\ref{sec:pconf}). For calibrated error bars, $u$ is uniformly distributed over the test catalogs. Figure~\ref{fig:pit} shows the histograms of $u$ for the six parameters in the three suites. A $\chi^2$ test of the ten bin counts against the uniform expectation gives $p>0.05$ in 13 of the 18 combinations. The five exceptions are $A_{\mathrm{SN2}}$, $A_{\mathrm{AGN1}}$ and $A_{\mathrm{AGN2}}$ in SIMBA and $\Omega_{\mathrm m}$ and $A_{\mathrm{AGN2}}$ in Astrid. The largest departure occurs for $A_{\mathrm{AGN2}}$ in SIMBA, where 37 of the 200 true values lie above the 90th percentile of their predictive distributions, against 20 expected. Over the 18 combinations, the 68\% and 90\% intervals contain 60\% to 78\% and 83\% to 97\% of the true values.